\documentclass{iopart}
\usepackage{graphicx}
\usepackage{amsfonts}
\usepackage{amssymb}

\usepackage{psfrag}

\def\be{\begin{equation}}
\def\ee{\end{equation}}
\def\ba{\begin{eqnarray}}
\def\ea{\end{eqnarray}}

\def\ket#1{|{#1}\rangle}
\def\bra#1{\langle{#1}|}

\begin{document}
\title{Numerical study of the critical phases
  of the frustrated $\mathbb{Z}(5)$ model}

\author{Christophe Chatelain}
\address{
Groupe de Physique Statistique,
D\'epartement P2M,
Institut Jean Lamour (CNRS UMR 7198),
Universit\'e de Lorraine, France}

\ead{christophe.chatelain@univ-lorraine.fr}

\date{\today}

\begin{abstract}
The phase diagram of the $\mathbb{Z}(5)$ spin model is studied
numerically on the square lattice by means of the {\sl Density Matrix
Renormalization Group}. In the regime where the two nearest-neighbor
couplings have opposite signs, a critical phase, not observed in
earlier Monte Carlo simulations, is identified. The new phase diagram
is in agreement with predictions made by M. den Nijs [{\sl Phys. Rev. B}
{\bf 31}, 266 (1985)] but for the $\mathbb{Z}(7)$ model rather than the
$\mathbb{Z}(5)$ one. All critical phases are shown to be compatible
with a central charge $c=1$. The magnetization scaling dimension displays
however a different behavior in the different critical phases.
\end{abstract}


\section{Introduction}
Frustration usually induces more complex, and therefore interesting,
new physics. In conjunction with disorder, like in glasses for example,
frustration leads, at sufficiently low temperature, to a complete freezing
of the system over accessible time scales. In pure systems, in
particular in magnetic systems, an increase of the degeneracy of the
ground state is observed when frustration is introduced. In some cases,
the ground state may even become infinitely degenerated,
leading to a finite entropy per site at zero temperature.
When thermally activated defects are interacting, the low-temperature
phase may be critical and bounded by a Berezinskii-Kosterlitz-Thouless
(BKT) topological phase transition~\cite{Berezinski,Kosterlitz}. Such
a situation will be considered in this paper.
\\

Frustration can be introduced mainly in three different ways in a
lattice spin model. First, a fraction of the interactions between spins
can be changed in order to favor locally different ground states.
A fraction of anti-ferromagnetic exchange couplings can be introduced
in a ferromagnetic spin model for example. An interesting example is
the fully-frustrated Ising model for which an odd number of exchange
couplings are anti-ferromagnetic in each plaquette of a square
lattice~\cite{Villain}. Consequently, the ferromagnetic phase is
unstable at any finite temperature. The anti-ferromagnetic Ising
model on a triangular lattice also belongs to this class of
fully-frustrated models. The latter does not display any anti-ferromagnetic
order at finite temperature but a new BKT topological transition driven
by the magnetic field appears in the limit of zero temperature~\cite{Blote}.
Similarly, Potts~\cite{Foster} and XY~\cite{Villain2}
fully-frustrated models were considered.
Even simpler two-dimensional models where horizontal bonds are
ferromagnetic while vertical ones are anti-ferromagnetic display
interesting new phase diagrams. The 3-state Potts model on a square
lattice with horizontal and vertical couplings, resp. $J_h$ and $J_v$,
such that $J_v.J_h<0$ undergoes a BKT transition too~\cite{Quartin}.
In three dimensions, Ding {\sl et al.} studied a mixed Potts model
with Hamiltonian~\cite{Ding}
\begin{equation}
  -\beta H=K\sum_{(i,j),z} \delta_{\sigma_{i,z},\sigma_{j,z}}
  -K\sum_{i,z} \delta_{\sigma_{i,z},\sigma_{i,z+1}}
\end{equation}
i.e. anti-ferromagnetic interactions in the $z$-direction and ferromagnetic
ones in the two other directions. In contrast to the 2D case, they found,
using Monte Carlo simulations, a second-order phase transition in the
universality class of the $O(n)$ model with $n=q-1$ and, when $q\ge 4$,
a discontinuous phase transition to a phase with a different type of
long-range order.
\\

The second manner to induce frustration in lattice spin models is to add
next-nearest neighbor interactions. The most celebrated model of this
kind is the ANNNI model~\cite{Selke}. In 2D, its phase diagram displays
a critical phase between the ferromagnetic and paramagnetic phases.
The three phases are separated by two BKT transitions. In 3D,
a so-called Lifshitz multicritical point lies at the meeting of the
two transition lines. Similarly, a 2D 3-state Potts
model with competing interactions between nearest and next-to-nearest
neighbors was studied by den Nijs~\cite{Nijs2}. The phase diagram
displays a critical phase, where the ground state and the thermally
activated defects can be mapped onto a 8-vertex model. The critical
exponents were inferred from this mapping.
\\

The third and less studied manner to induce frustration is to introduce
two different interactions between each pair of nearest neighbors of
the lattice. Lee and Grinstein considered the generalized XY model
with Hamiltonian~\cite{LeeGrinstein}:
\begin{equation}
  -\beta H=\sum_{(i,j)} \big[J_1\cos(\theta_i-\theta_j)
    +J_2\cos(2\theta_i-2\theta_j)\big]
\end{equation}
where $\theta_i\in [0;2\pi[$.
Besides the vortices giving rise to the usual BKT transition, they found
half-integer vortices and string excitations. The phase diagram displays
lines of BKT transitions and lines of Ising transitions.
Dian {\sl et al.} considered the Lee-Grinstein model with $J_2<0$ and
found an additional spin-ice phase~\cite{Dian}. Poderoso {\sl et al.}
extended this model to~\cite{Poderoso}
\begin{equation}
  -\beta H=\sum_{(i,j)} \big[\Delta\cos(\theta_i-\theta_j)
    +(1-\Delta)\cos(q\theta_i-q\theta_j)\big].
\end{equation}
It was later shown that for $q=3$ the Ising line is replaced by a line
in the universality class of the 3-state Potts model~\cite{Canova}.
In this paper, the $\mathbb{Z}(q)$ model, a discretized version of the
Lee-Grinstein model whose Hamiltonian is~\cite{Jose}
\begin{equation}
  -\beta H=\sum_{(i,j)} \Big[J_1\cos {2\pi\over q}(\sigma_i-\sigma_j)
    +J_2\cos {4\pi\over q}(\sigma_i-\sigma_j)\Big].
\end{equation}
where $\sigma_i\in\{0\ldots q-1\}$, is considered.
The case $q\le 4$ is equivalent to the usual
$q$-state Potts model, for which the phase diagram displays a single
transition point separating the ferromagnetic and paramagnetic phases.
The case $q\ge 5$ allows for the existence of spin-wave excitations
and, as a consequence, of a critical phase between the ferromagnetic
and paramagnetic phase. The $\mathbb{Z}_q$ model also allows for
frustrated couplings, which lead to a richer phase diagram with new
critical phases. den Nijs discussed this phase diagram and
proposed several scenarios for the sequence of transitions undergone
by the model~\cite{Nijs}. The phase diagram of
the $\mathbb{Z}(5)$ model (including the anti-ferromagnetic regime
$J_1<0$) was shortly after determined numerically by Monte Carlo
simulations~\cite{Baltar}. While the ferromagnetic regime $J_1>0$
is well reproduced by the numerical data,
discrepancies with den Nijs' predictions were observed in the
anti-ferromagnetic regime. The accuracy was further improved
by later numerical studies of the ferromagnetic regime~\cite{Rouidi},
but, as far as we are aware, the anti-ferromagnetic regime
has not been studied in the last three decades. A much more
accurate test of den Nijs' predictions is now possible.
\\

In the following, a numerical study of the phase diagram of the
$\mathbb{Z}(5)$ model is presented. In section 2, known results
concerning this model are briefly summarized and den Nijs'
interpretation of the mechanism of the transitions is presented.
In section 3, different observables, numerically computed, are
discussed to determine the phase diagram. Discrepancies are
found with earlier Monte Carlo simulations. The nature of the phase
transitions in the regime $J_1<0$ is then studied by considering the
lattice size behavior at some particular values of the coupling $J_1$.
Helicity modulus and entanglement entropy are considered.
Finally, the magnetization scaling dimension is estimated by a
log-log fit of spin-spin autocorrelation functions.
Conclusions follow.

\section{The $\mathbb{Z}(5)$ model and its phase diagram}
\label{Intro}
We consider the lattice spin model defined by the following Hamiltonian:
\begin{equation}\fl
  -\beta H=\sum_{(i,j)}\Big[J_1\cos{2\pi\over q}(\sigma_i-\sigma_j)
  +J_2\cos{4\pi\over q}(\sigma_i-\sigma_j)\Big]+h\sum_i
  \cos{2\pi\over q}\sigma_i
\end{equation}
with $q=5$. The spin $\sigma_i$ lies on the $i$-th node of a square
lattice. The first two terms couple the pairs $(i,j)$ of nearest-neighbors
on the lattice. Both favor a ferromagnetic ordering when $J_1$ and $J_2$ are
positive. Frustration is induced if $J_1.J_2<0$. The last term is
the Zeeman Hamiltonian coupling the spins with an external magnetic field.
While the other terms are invariant under any cyclic permutation of the
$q$ states, i.e. under the transformations of the group ${\mathbb Z}(q)$,
the Zeeman Hamiltonian breaks this symmetry. The definition of
magnetization $M=\sum_i \langle\cos{2\pi\over q}\sigma_i\rangle$
is immediately implied by the Zeeman Hamiltonian. In the following,
we will restrict ourselves to the case $h=0$ in the bulk of the system.
Magnetic fields will only be applied at the boundaries.
\\

When $J_2=0$, the 5-state clock model is recovered~\cite{Potts}.
Along this line, the ferromagnetic and paramagnetic phases are separated
by a thin critical phase, bounded by two BKT phase transitions.
When $J_1=J_2$, one can check that the Hamiltonian reduces to
\begin{equation}
  -\beta H=J\sum_{(i,j)} \Big[2\delta_{\sigma_i,\sigma_j}-{1\over 2}
  (1-\delta_{\sigma_i,\sigma_j})\Big]=J\sum_{(i,j)} \Big[{5\over 2}
    \delta_{\sigma_i,\sigma_j}-{1\over 2}\Big]
\end{equation}
i.e. the 5-state Potts model~\cite{Potts}. Therefore, the system
undergoes a weak first-order phase transition for $J_1=J_2={2\over 5}
\ln(1+\sqrt 5)\simeq 0.470$. The self-duality of the Potts model
can be extended to the case $J_1\ne J_2$~\cite{Alcaraz}. Introducing
the two parameters
\begin{equation}
  x_1=e^{\mu_1J_1+\mu_2J_2},\hskip 1truecm
  x_2=e^{\mu_1J_2+\mu_2J_1}
  \label{Defx}
\end{equation}
where
\begin{equation}
  \mu_1=\cos{2\pi\over 5}-1={\sqrt 5-5\over 4},\hskip 1truecm
  \mu_2=\cos{4\pi\over 5}-1=-{\sqrt 5+5\over 4}
\end{equation}
the line of self-duality is given by~\cite{Rouidi}
\begin{equation}
  x_1+x_2={\sqrt{5}-1\over 2}.
\end{equation}
The system undergoes a single phase transition along a finite portion
of this self-dual line that is delimited by the two Fateev-Zamolodchikov
points where the model is integrable by algebraic Bethe ansatz~\cite{Fateev}.
The transition point of the Potts model ($J_1=J_2$) belongs to this portion
of the self-dual line.
Between the two Fateev-Zamolodchikov points, the transition is of
first-order while it is of second order at these points. Outside of
this finite portion of the self-dual line, the system undergoes two
BKT transitions and the self-dual line is in the critical phase.
The clock model $J_2=0$ corresponds to this situation. A second
branch of the self-dual line is in the region $J_1<0$ and $J_2>0$.
According to earlier Monte Carlo simulations, the line lies in a critical
phase~\cite{Baltar}.

\begin{figure}[ht]
\centering
\psfrag{r}[Bc][Bc][1][1]{$r$}
\psfrag{E/J1}[Bc][Bc][1][0]{$-\beta{\cal H}/J_1$}
\centerline{
\includegraphics[width=7cm]{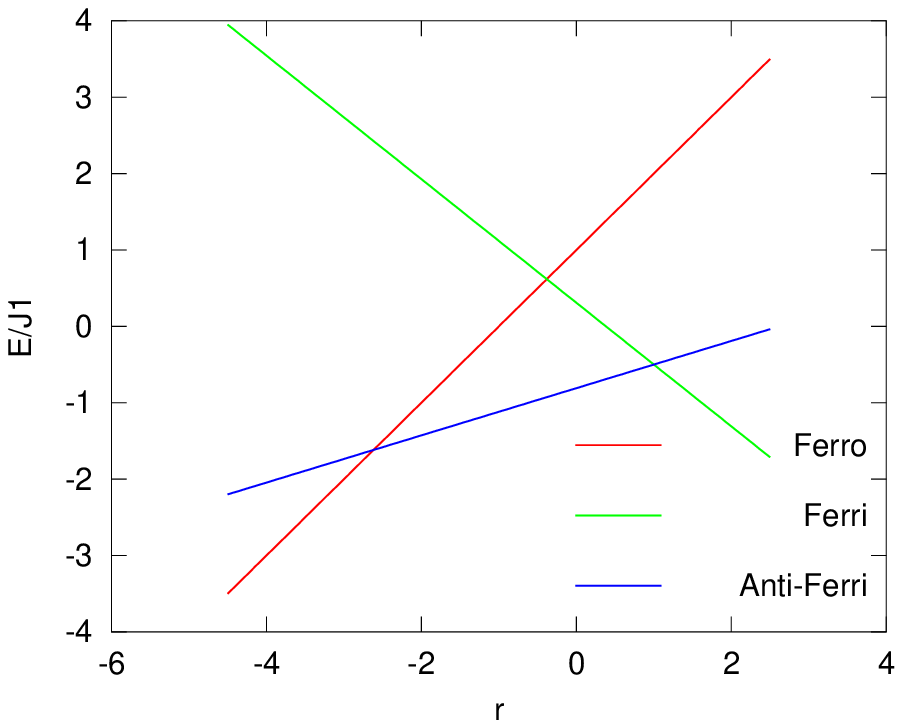}\quad
\includegraphics[width=6cm]{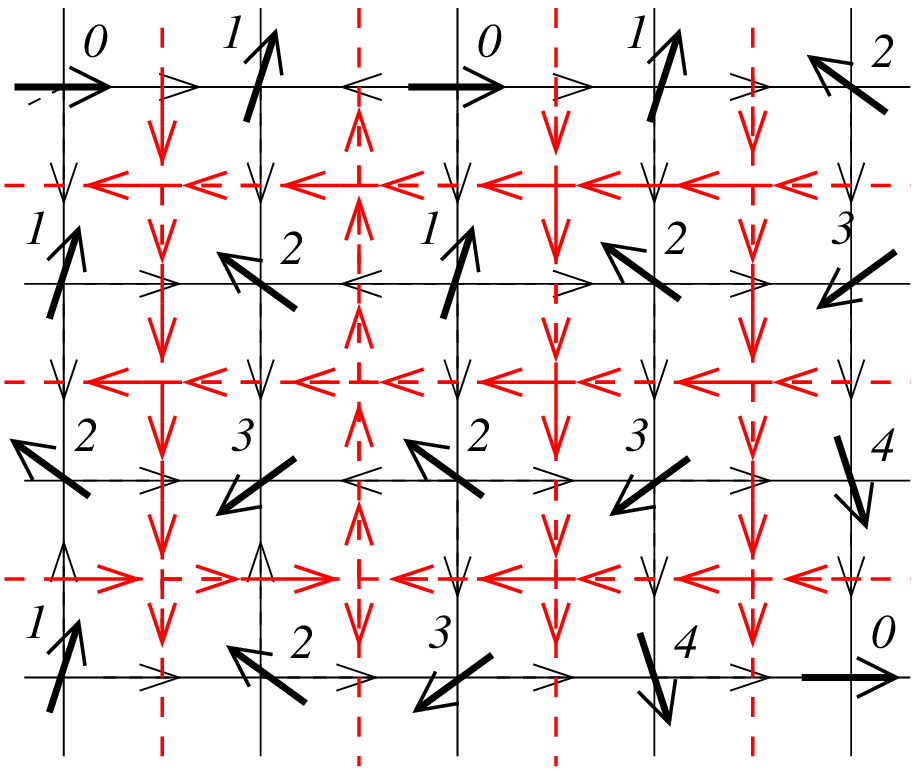}
}\caption{On the left, reduced values $-\beta H/J_1$ of the energy of a
bond between two spins versus $r=J_2/J_1$. The different curves corresponds
to the three possible relative states $0$, 1 or 2 of the spins
$|\sigma'-\sigma|\ {\rm mod}\ q$. On the right, example of a spin
configuration in the ferrimagnetic phase in the limit $J_1\rightarrow
\pm\infty$. The spins are represented as bold black arrows. Light black
arrows are the bond variables $d_{ij}$ and the red arrows at the center
of each plaquette correspond to the equivalent six-vertex ice model.
}
\label{FigT0}
\end{figure}

The phase diagram is readily obtained in the two limits $J_1\rightarrow
\pm \infty$. As discussed for instance in Ref.~\cite{Baltar}, the energy
$\beta H$ of the bond connecting the two spins $\sigma$ and $\sigma'$ takes
only three possible values~:
   \be -\beta H=\left\{
     \begin{array}{c c}
       J_1+J_2 & (\sigma=\sigma') \\
       aJ_1+bJ_2 &  (\sigma=\sigma'\pm 1\ {\rm mod}\ q)\\
       bJ_1+aJ_2 &  (\sigma=\sigma'\pm 2\ {\rm mod}\ q)\\
     \end{array}
   \right.
   \ee
where $a=\cos{2\pi\over 5}$ and $b=\cos{4\pi\over 5}$. Using a magnetic
language, the first case corresponds to a ferromagnetic state, the second to
a ferrimagnetic one and the third to an anti-ferrimagnetic state.
On figure~\ref{FigT0}, the different energies $-\beta H/J_1$ are represented
with respect to $r=J_2/J_1$. In the limit $J_1\rightarrow +\infty$, the
ground state is given by the maximum of $-\beta H/J_1$. The ground state
is therefore ferromagnetic for $r>r_*=(1-a)/(b-1)\simeq -0.382$
and ferrimagnetic otherwise. In the limit $J_1\rightarrow -\infty$, the
ground state is given by the minimum of $-\beta H/J_1$. For $r>1$, the
ground state is ferrimagnetic. For $r<1/r_*=(b-1)/(1-a)\simeq -2.618$, the
ferromagnetic state is stabilized by the positive coupling $J_2$ which
is larger than $|J_1|$. Between these two phases, i.e. for $r\in [1/r^*;1]$,
the ground state is anti-ferrimagnetic.
\\

In the absence of external magnetic field, the ferromagnetic phase
is $q$-fold degenerated. In contrast, the ferrimagnetic and
anti-ferrimagnetic ground states are infinitely degenerated. Consider
for example the spin configuration in the ferrimagnetic phase presented
on figure~\ref{FigT0}. The spins are represented as black bold arrows
located on the sites of the square lattice and making an angle
$\theta_i={2\pi\over 5}\sigma_i$ with the $x$-axis. The Potts
states $\sigma_i$ associated to each spin orientation is written
by the side of the arrow. In the ferrimagnetic ground state,
the difference between two neighboring spins $\sigma_i$ and $\sigma_j$
is expected to be $\sigma_i-\sigma_j=\pm 1\ {\rm mod}\ q$.
The infinite degeneracy of the ground state
comes from the fact that, if three spins of a square plaquette are already
fixed, there still exists two possibilities for the fourth one if its
two neighbors are in the same state.
\\

It is useful to consider the bond variables $d_{ij}=\sigma_i-\sigma_j
\ {\rm mod}\ q$. As mentioned above, $d_{ij}=\pm 1$ in the ground state of
the ferrimagnetic phase. By construction, the $d_{ij}$ are constrained by the
condition $\sum_{(i,j)\in\square_\alpha} d_{ij}=0\ {\rm mod}\ q$ around each
plaquette $\square_\alpha$ of the lattice. In the ferrimagnetic ground state,
this condition can only be satisfied
if, around each plaquette, $d_{ij}=+1$ for two of the four bonds and $d_{ij}
=-1$ for the two others. The circulation $\sum_{(i,j)} d_{ij}$ vanishes for
all plaquettes. On figure~\ref{FigT0}, the bond variables $d_{ij}=\sigma_i
-\sigma_j\ {\rm mod}\ q$ between each neighboring spins $\sigma_i$ and
$\sigma_j$ are represented as light black arrows at the center of the lattice
bond joining the two spins. As a convention, the arrow points towards the
spin which is larger by 1 (modulo $q$) than the other one. In the
ferrimagnetic (as well as anti-ferrimagnetic) ground state, there are
six possible arrow configurations compatible with the constrain of
a vanishing circulation. In the thermodynamic limit, there are an
infinite number of ways of piling-up these plaquettes, and therefore
an infinite number of possible spin configurations in the ground state.
As pointed out by den Nijs~\cite{Nijs}, the configuration of these arrows
can be interpreted as the height differences $h_i-h_j$ of a restricted
Solid-On-Solid model (RSOS). At finite $J_1$, the phase is critical
due to the proliferation of massless spin wave excitations.
The same conclusions can be drawn for the anti-ferrimagnetic phase,
the only difference being that the difference between two neighboring
spins $\sigma_i$ and $\sigma_j$ is expected to be $d_{ij}=\pm 2$.
Following den Nijs, double arrows will be drawn on the bond joining
the two spins $\sigma_i$ and $\sigma_j$.
\\

The arrow configurations in the limit $|J_1|\rightarrow +\infty$
can mapped onto a 6-vertex model. Such a mapping, mentioned by
den Nijs in the case of the $\mathbb{Z}(q)$ model, was recently
described for the ground state of the Lee-Grinstein model~\cite{Dian}.
Consider again the spin configuration
in the ferrimagnetic ground state depicted on figure~\ref{FigT0}.
The dual lattice is represented as red (or light gray) dashed lines.
The sites of this dual lattice lie at the center of the plaquettes
of the original square lattice. Each one of the four bonds emerging
from any site of the dual lattice crosses one of the four bonds of
the plaquette. Above, these bonds of the original lattice were
given an orientation according to the sign of the difference $d_{ij}$
of the two spins at their edges. In the following, an orientation
will also be given to the bonds of the dual lattice. Consider one site
of the dual lattice. Follow one of the bonds emerging from this site.
Eventually, a bond of the original lattice will be crossed. Determine
its orientation. If the latter corresponds to the trigonometric
orientation around the plaquette, an arrow pointing
outwards is placed on the bond of the dual lattice. In the other case,
the arrow is inwards. As can be observed on figure~\ref{FigT0}, this
set of arrows on the dual lattice, denoted $d_{\alpha\beta}^*$, corresponds
to a configuration of the 6 vertex model. In particular, as a consequence
of the constrain of vanishing circulation $\sum_{(i,j)\in\square_\alpha} d_{ij}=0$,
the ice rule is satisfied: two inwards and two outwards arrows meet
at each site of the dual lattice. While the bond variables $d_{ij}$
are irrotationnal, their duals $d_{\alpha\beta}^*$ are divergenceless.
Since $d_{ij}\rightarrow d_{\alpha\beta}^*$ is a one-to-one mapping,
the degeneracy of the ground state is the same for the $\mathbb{Z}(q)$
model and the 6-vertex model at the ice point, up to a factor $q$
corresponding to the state of the first spin. Therefore, the entropy
per site takes the exact value ${3\over 2}\ln {4\over 3}$
\cite{Lieb}. The same discussion applies to
the anti-ferrimagnetic phase, apart from the fact that $d_{ij}=\pm 2$
and therefore $d_{\alpha\beta}^*=\pm 2$ too.
\\

As confirmed by Monte Carlo simulations, the model undergoes a phase
transition from the (anti)-ferrimagnetic phase to the paramagnetic one
at finite $J_1$~\cite{Baltar}. The question of the mechanism of
these phase transitions is not settled.
The possible scenarios discussed by den Nijs are summarized in the following.
In the ferrimagnetic phase (or floating solid phase), consisting only of
single arrows in the limit $|J_1|\rightarrow +\infty$, local defects
are excited at finite $J_1$. On the square lattice, they consist of
one bond with a bond variable $d_{ij}=\pm 2$. Such a defect should
occur with a probability $\sim e^{J_1(1-r)(b-a)}$. The constrain of a
vanishing circulation, $\sum_{(i,j)\in\square_\alpha} d_{ij}=0$, cannot be
satisfied with one bond $d_{ij}=\pm 2$ and three $d_{ij}=\pm 1$.
However, the $\mathbb{Z}(q)$ symmetry allows to consider plaquettes
with $\sum_{(i,j)\in\square_\alpha} d_{ij}=\pm q$ which is obviously satisfied
with one bond $d_{ij}=+2$ and three $d_{ij}=+1$ (or one $-2$ and three $-1$).
These excited plaquettes, called vertices by den Nijs, are topological
defects with charges $+q$ and $-q$, depending on their vorticity. In
the SOS model, they correspond to screw dislocations. The excited bond
$d_{ij}=\pm 2$ being shared between two plaquettes, two $\pm q$
defects are simultaneously formed. Defects consisting of plaquettes
with two excited bonds, either identical, as $\{+1,-1,+2,-2\}$ and
$\{+1,-1,0,0\}$, or different $\{-1,-1,+2,0\}$ may also appear but
with a smaller probability~\footnote{respectively $e^{2J_1(1-r)(b-a)}$,
$e^{2J_1[1-a+(1-b)r]}$ and $e^{J_1[(1+b-2a)+(1+a-2b)r]}$.}.
All of them have a vanishing circulation (or vorticity). In contrast
to the $\pm q$-vertex which are found only by pairs, these $0$-vorticity
defects exist only as strings. The plaquettes forming these strings
are connected by their excited bonds. The strings either form loops or
are terminated by two $\pm q$-vertices at each edge. The latter is a composite
object called a $2q$-vortex by den Nijs. In the anti-ferrimagnetic phase,
for which all bonds variables are $d_{ij}=\pm 2$ in the ground state, local
defects correspond to pairs of plaquettes with one bond with
$d_{ij}=\pm 1$. These defects have a circulation $\pm 2q$.
As in the ferrimagnetic phase, defects consisting of plaquettes with
two excited bonds, for example $d_{ij}\in\{+2,-2,+1,-1\}$ or $d_{ij}\in
\{+2,-2,0,0\}$, and therefore with a vanishing circulation (or vorticity),
may also appear but with a smaller probability. They form either loops
or strings connecting two $\pm 2q$-vertices. The latter is a $4q$-vortex.
\\

As in the celebrated XY model, a logarithmically decaying interaction
binds the vortices by pair in the critical phases. Several scenarios are
described by den Nijs: if the string tension binding the vertices is stronger
than the interaction between vortices, the system will undergo a BKT
transition (case a). In the high-temperature paramagnetic phase, vortices
are free. On the other hand, the string tension may be weaker than the
interaction between vortices. Then, $nq$-vortices split
into free vertices or $mq$-vortices with $m<n$, i.e. a smaller
topological charge. If the temperature is already higher
than the BKT-temperature of these $mq$-vortices, they will never form
pairs and the string melting transition brings the system directly into
the paramagnetic phase (case b). If it is not the case, a BKT transition
will occur at high temperature and string melting is only a cross-over
(case c). In all cases, free vortices may still exist in the paramagnetic
phase. At much higher temperature, the string binding the vertices will
melt and the vortices will split into free vertices. However, no phase
transition will be undergone (case d).
\\

As already mentioned in the introduction, the half-plane $J_1>0$ of the
phase diagram is well-known and accurate numerical results can be found in
the literature. In contrast, the case $J_1<0$ was not studied in the last
three decades and is not known with a good accuracy. Up to now, den Nijs'
proposed phase diagram is not confirmed by Monte Carlo simulations.

\section{Numerical results}
We studied numerically the $\mathbb{Z}(5)$ model using the algorithm
{\sl Density Matrix Renormalization Group}~\cite{White,Schollwock} (DMRG).
The ground state of the transfer matrix is expressed as a matrix product
state of two blocks, left and right. In the determination of the phase
diagram, these two blocks were allowed to grow up to $325$ states. This
number of states was increased for the small subset of points where the
diagonalization routine failed to converge. The ground state, first
obtained by the Infinite-Size DMRG algorithm, was refined
using six sweeps of the Finite-Size DMRG algorithm. The phase diagram was
constructed by performing calculations for the lattice size $L=128$
for 60 different values of $J_1$ and 60 of $J_2$. Additional
simulations for lattice sizes $L=32$, 48, 64 and 96 at some potentially
interesting values of $J_1$ were then performed.

\subsection{DMRG estimation of the phase diagram}
\label{Sec1}

\begin{figure}[ht]
\psfrag{r=J2/J1}[Bc][Bc][1][1]{$r$}
\psfrag{T=1/J1}[Bc][Bc][1][0]{$T$}
\centerline{
  \includegraphics[width=8.34cm,height=6.63cm]{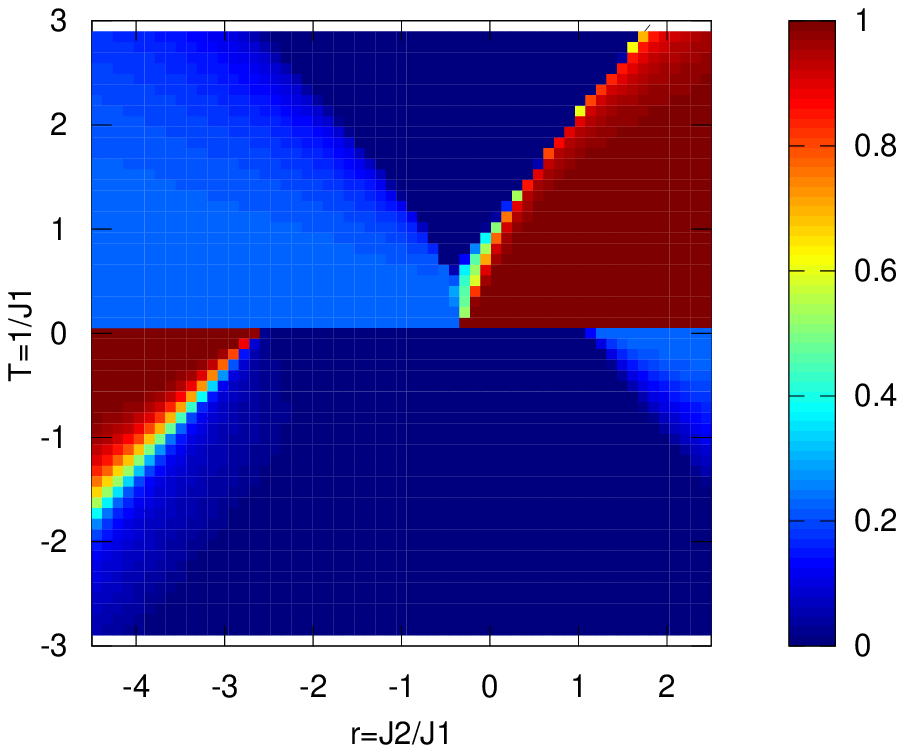}
  \includegraphics[width=6.63cm,height=6.63cm]{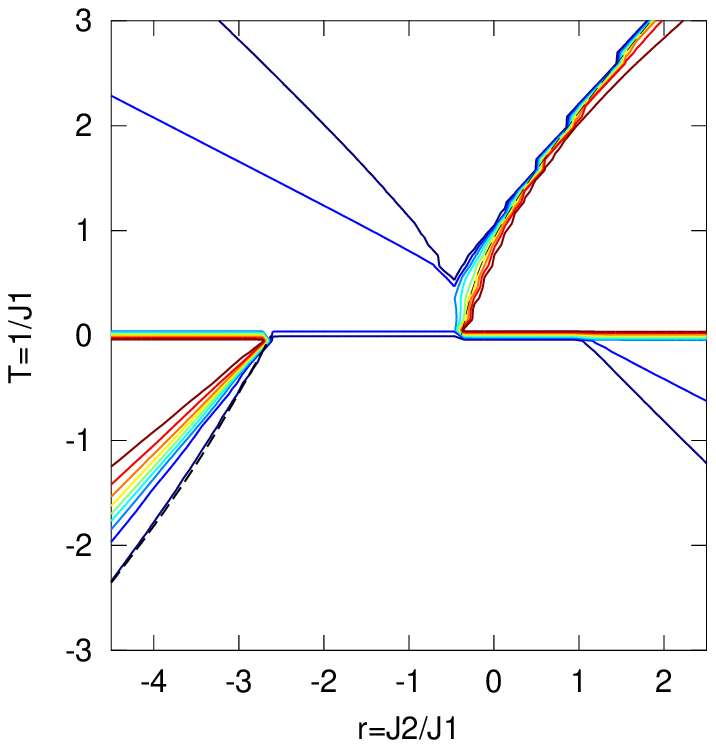}}
\caption{Magnetization $\langle\cos{2\pi\over 5}\sigma_{L/2}\rangle$
  versus $r=J_2/J_1$ and $T=1/J_1$ for a lattice size $L=128$ obtained
  by DMRG with $325$ states and 6 sweeps. On the right, contour plot
  of the same data. The different contour lines correspond to values
  of magnetization equal to multiples of $0.1$. The dashed line is
  the self-dual line of the model. In the lower half-plane, it
  almost coincides with the contour line $m=0.1$.}
\label{Fig1}
\end{figure}

The two ferromagnetic phases are easily localized using the local
magnetization $m=\langle\cos{2\pi\over 5}\sigma_{L/2}\rangle$ measured
at the center of the chain (Figure~\ref{Fig1}). To break the
$\mathbb{Z}(5)$-symmetry and induce a non-vanishing spontaneous
magnetization in these ferromagnetic phases, magnetic fields were
introduced at the boundaries of the system with the Hamiltonian
\begin{equation}
  -\beta H_{\rm BC}=\cos{2\pi\over q}\sigma_1
  +\cos{2\pi\over q}\sigma_L.\label{BoundaryH}
\end{equation}
  The convergence of the DMRG algorithm is also improved by these
boundary magnetic fields.
As expected, the ferromagnetic phases extend over the interval
$r=J_2/J_1\in ]-\infty;-2.618]$ when $J_1\rightarrow -\infty$
and $[-0.382;\infty[$ when $J_1\rightarrow +\infty$. The location
of the phase boundaries are in good agreement with earlier Monte
Carlo simulations. In the contour plot of figure~\ref{Fig1}, one
can observe that the contour line of smallest magnetization is
close to the self-dual line, apart in the region between $r\simeq
-0.382$ and the first Fateev-Zamolodchikov point at $r=(\sqrt 5-1)/2
\simeq 0.618$ where there exists a critical phase between the
paramagnetic and ferromagnetic phases. In our numerical data,
the critical phase manifests itself as a broadening of the region
where magnetization displays a significant variation. Note that
for very negative values of $r$, i.e. in the region $r\lesssim -3$,
a region that has not been considered numerically before, a much
stronger broadening is observed. On figure~\ref{Fig1}, the phase
boundary does not appear sharp but spread over an interval of
values of $r$ growing with $|J_1|$. It is tempting to conjecture
that a critical phase also lies between the ferromagnetic and
paramagnetic phase in the lower half-plane. This point will be
discussed again in the following. Ferrimagnetic phases can also
be observed, and distinguished from the ferromagnetic phases,
since they display a small but non-vanishing spontaneous
magnetization. In the upper-half plane $J_1>0$, the phase boundary
with the paramagnetic phase is in fairly good agreement with
Ref.~\cite{Baltar}. In contrast, the agreement is quite poor
in the lower half-plane. Last, the anti-ferrimagnetic
phase is invisible to magnetization.

\begin{figure}[ht]
\psfrag{r=J2/J1}[Bc][Bc][1][1]{$r$}
\psfrag{T=1/J1}[Bc][Bc][1][0]{$T$}
\centerline{
  \includegraphics[width=8.34cm,height=6.63cm]{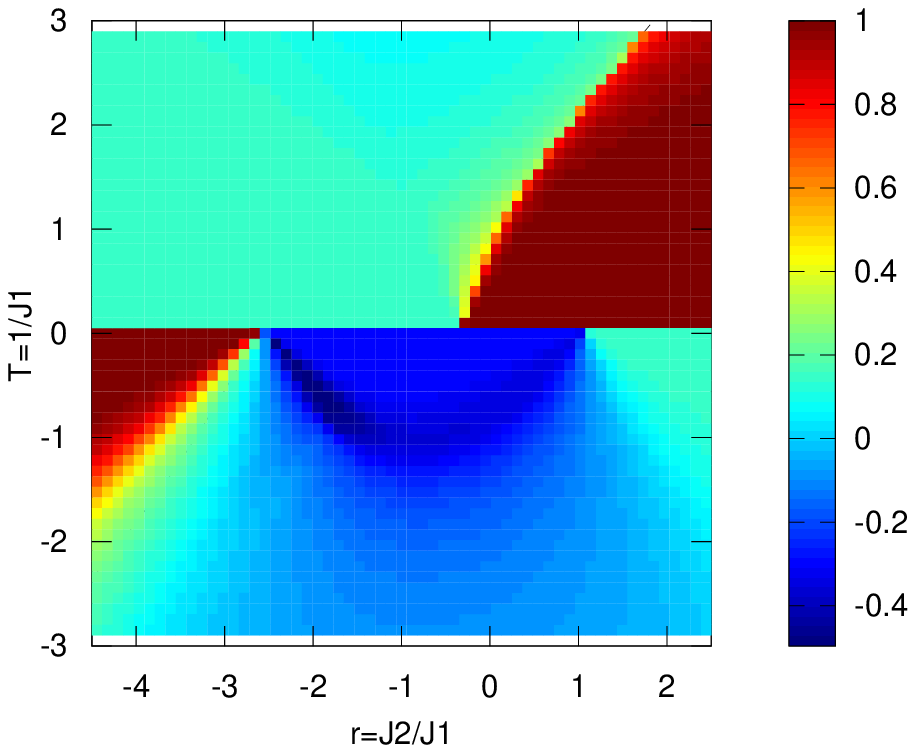}
  \includegraphics[width=6.63cm,height=6.63cm]{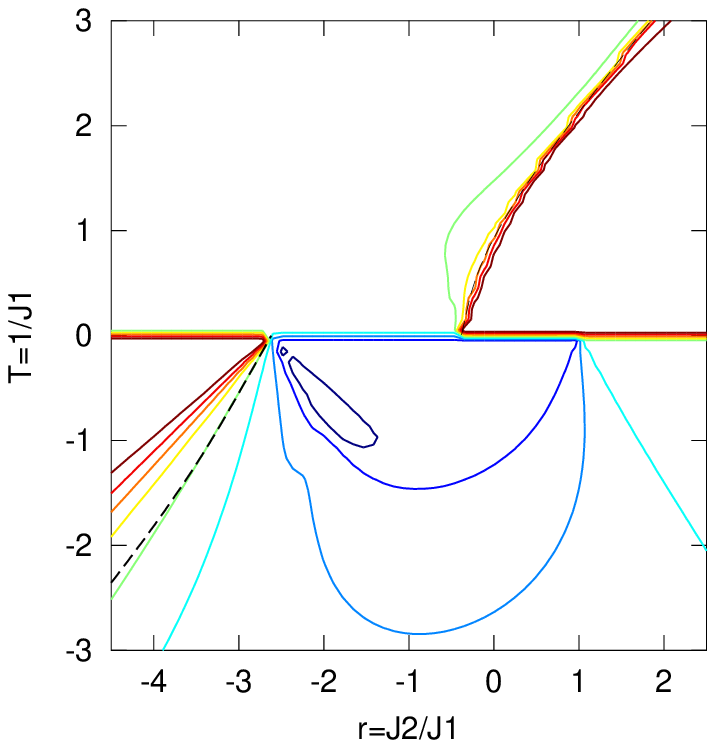}}
  \caption{Magnetization-magnetization autocorrelation function $A_1$
  versus $r=J_2/J_1$ and $T=1/J_1$ for a lattice size $L=128$ obtained
  by DMRG with $325$ states and 6 sweeps.
  On the right, contour plot of the same data. The contour lines
  correspond to values of $A_1$ which are multiples of $0.15$.
  The dashed line is the self-dual line of the model. In the lower
  half-plane, the self-dual line almost coincides with the (green)
  contour line $A_1=0.2$. The cyan contour line corresponds to
  a vanishing $A_1$. The contour line $A_1=-0.4$ mentioned in the
  text appears in blue and looks like a half-circle.}
\label{Fig5}
\end{figure}

Another quantity that can be computed to identify the magnetic order
in the system is the magnetization-magnetization autocorrelation
function
\begin{equation}
  A_1=\langle\cos{2\pi\over 5}\sigma_{L/2}(1)
  \cos{2\pi\over 5}\sigma_{L/2}(0)\rangle
\end{equation}
where the central local magnetization is measured at two successive
times $t$. In this context, the time direction corresponds to the
transverse direction of the transfer matrix. As can be seen on
figure~\ref{Fig5}, the two ferromagnetic phases appear distinctly,
as well as the anti-ferrimagnetic phase in the lower half-plane
$J_1<0$. The two ferrimagnetic phases are now hardly distinguishable from
the paramagnetic phase. It is however difficult to estimate the precise
location of the phase boundary of the anti-ferrimagnetic phase with the
paramagnetic one. Since $A_1$ vanishes in a purely paramagnetic phase
while $A_1=\cos{4\pi\over 5}\simeq -0.81$ in an anti-ferrimagnetic phase
in the limit $J_1\rightarrow -\infty$, this phase boundary may be estimated
as the location of the points where $A_1\simeq -0.4$. The corresponding
curve can be seen on the contour plot of figure~\ref{Fig5}. Of course,
the choice of the value $-0.4$ introduces some arbitrariness. As seen
on the figure, small changes to this value lead to a quite different
contour plot. Nevertheless, if the boundary of the anti-ferrimagnetic
phase follows one of these contour lines then this phase occupies only
a finite region of the phase diagram, in contrast to the infinite phase
suggested by Monte Carlo simulations. Moreover, it does not share any
common boundary
with the ferromagnetic phase, apart from the point $r\simeq -2.618$ when
$J_1\rightarrow -\infty$, as stated by Ref.~\cite{Baltar}.
This discrepancy is important: den Nijs' proposal was rejected by
Ref.~\cite{Baltar} because of the observation in the Monte Carlo
simulations of a common phase boundary between the ferromagnetic
phase and the anti-ferrimagnetic one in the half-plane $J_1<0$. The
absence of any common phase boundary, as observed in our DMRG
calculations, is in agreement with den Nijs' proposal.

Moreover, the critical phase between the ferromagnetic and paramagnetic
phases appear distinctly in yellow color ($A_1\simeq 0.4$). Figure~\ref{Fig5}
confirms that a critical phase lies at the boundary of the ferromagnetic
phase, not only in the upper half-plane $J_1>0$ but also in the lower one.
In the former, the BKT transition lines are known to merge at the two
Fateev-Zamolodchikov points, while in the latter, the BKT lines are observed
to meet only on the $T=0$-axis at $r\simeq -2.618$. Other merging points
may exist outside of the considered region, i.e. for $r<-4.5$.

\begin{figure}[ht]
\psfrag{r=J2/J1}[Bc][Bc][1][1]{$r$}
\psfrag{T=1/J1}[Bc][Bc][1][0]{$T$}
\centerline{
  \includegraphics[width=8.34cm,height=6.63cm]{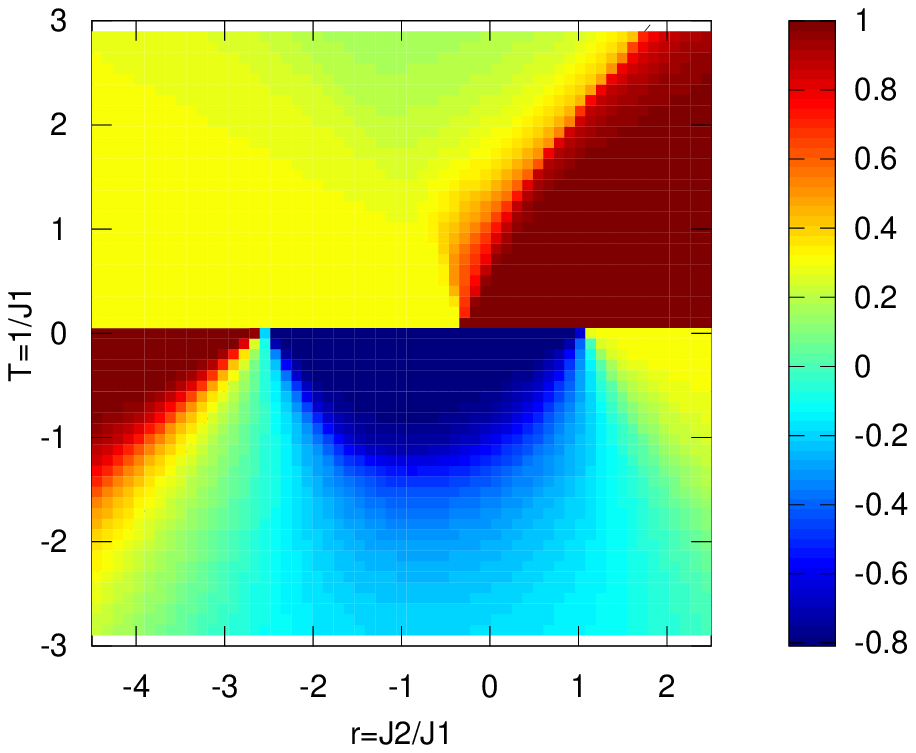}
  \includegraphics[width=6.63cm,height=6.63cm]{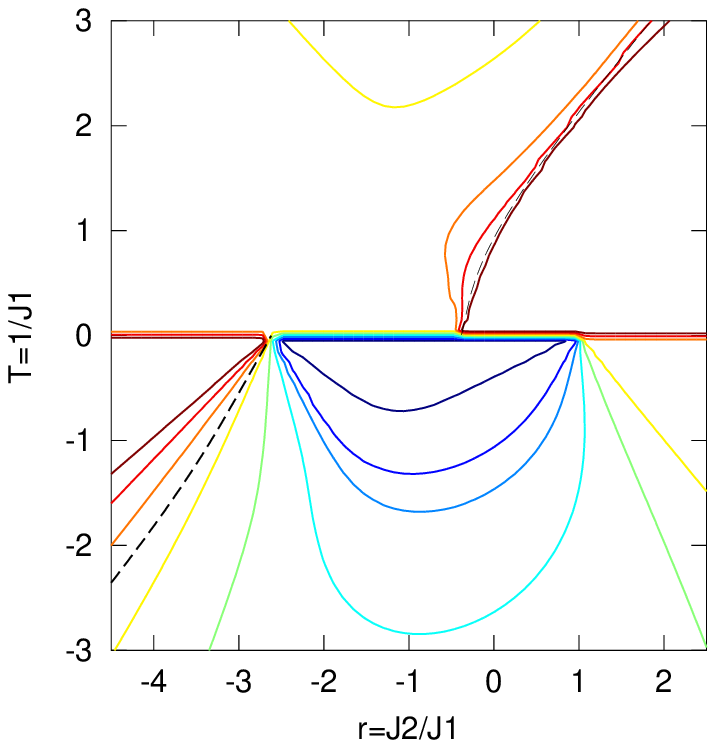}}
  \caption{Two-site correlation $e=\langle\cos{2\pi\over 5}(\sigma_{L/2}
  -\sigma_{L/2+1})\rangle$ versus $r=J_2/J_1$ and $T=1/J_1$ for a lattice
  size $L=128$ obtained by DMRG with $325$ states and 6 sweeps.
  On the right, contour plot of the same data. The contour lines
  correspond to values of the two-site correlation which are multiples
  of $0.2$. The dashed line is the self-dual line of the model.
  In the lower half-plane, it is located between the contour lines
  $e=0.2$ (yellow) and $0.4$ (orange). The green contour line corresponds
  to a vanishing two-site correlation.
  }
\label{Fig7}
\end{figure}

A closely-related observable allowing to distinguish the different phases
is the nearest-neighbor spin-spin correlation
\begin{equation}
  e=\langle\cos{2\pi\over q}(\sigma_{L/2}-\sigma_{L/2+1})\rangle.
\end{equation}
again measured at the center of the chain. In the limit $|J_1|\rightarrow
+\infty$, $e$ is expected to take the values 1 in the ferromagnetic
phases, $\cos{2\pi\over 5}\simeq 0.309$ in the ferrimagnetic phases,
$\cos{4\pi\over 5}\simeq -0.81$ in the anti-ferrimagnetic phase and to
vanish in the paramagnetic phase. On figure~\ref{Fig7}, the two
ferromagnetic phases appear as red, i.e. $e\simeq 1$, and at the
expected location. In the two phases of partial magnetic order previously
observed, the two-site correlation $e$ is compatible with $e\simeq 0.309$,
in agreement with the interpretation of these phases as ferrimagnetic.
The anti-ferrimagnetic phase is now better resolved than with
$A_1$. The different contour lines are indeed closer to each other.
As a consequence, defining the phase boundary with the paramagnetic
phase as the contour line $e=e_{PB}=0.4$ leads to a more stable
result upon small variations of $e_{PB}$ than with $A_1$.
The discrepancy with earlier Monte Carlo simulations is confirmed.
\\

\begin{figure}[ht]
\psfrag{r=J2/J1}[Bc][Bc][1][1]{$r$}
\psfrag{T=1/J1}[Bc][Bc][1][0]{$T$}
\centerline{
  \includegraphics[width=8.34cm,height=6.63cm]{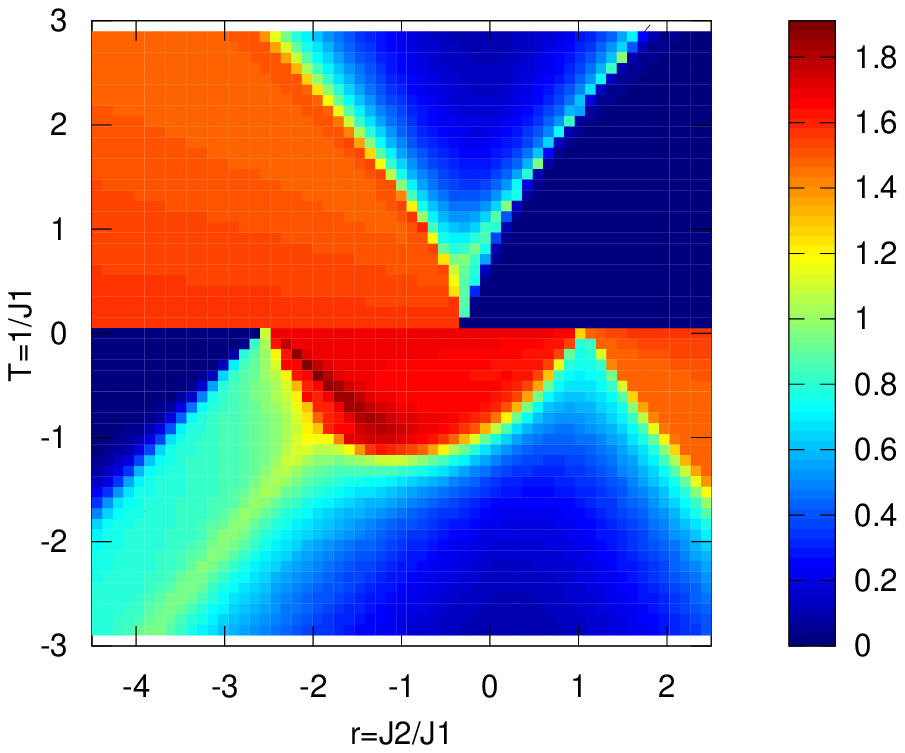}
  \includegraphics[width=6.63cm,height=6.63cm]{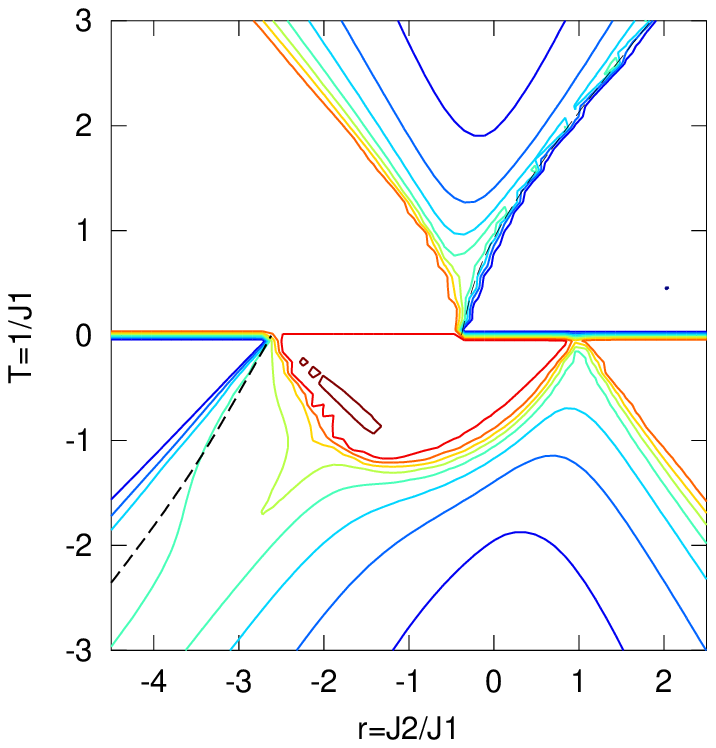}}
\caption{Entanglement entropy $S(L/2)$ versus $r=J_2/J_1$ and $T=1/J_1$
  for a lattice size $L=128$ obtained by DMRG with $325$ states and 6 sweeps.
  On the right, contour plot of the same data. The contour lines correspond
  to values of the entanglement entropy which are multiples of $0.2$.
  The dashed line is the self-dual line of the model. Close to the
  $T=0$ axis, the self-dual line almost merges with the $S=0.8$
  contour line.}
\label{Fig3}
\end{figure}

\def\trace{\mathop{\rm Tr}\nolimits}

Finally, probably the most stable locations of the phase boundaries
are given by the entanglement entropy. We will restrict ourselves to
the discussion of the entanglement entropy in the ground state
of one half of the system with the second half. This
quantity will be denoted $S(L/2)$, where $L$ is the lattice size.
The Hilbert space
being decomposed as a tensor product of the left and a right
Hilbert spaces, i.e. ${\cal H}={\cal H}_L\otimes{\cal H}_R$, the
entanglement entropy is defined as the average of the logarithm
of the reduced density matrix of one of the halves~\cite{Amico}:
\begin{equation}
  S(L/2)=-\trace_{{\cal H}_L}\rho_L\ln\rho_L
\end{equation}
where
\begin{equation}
  \rho_L=\trace_{{\cal H}_R} \rho
  =\trace_{{\cal H}_R} \ket{\psi_0}\bra{\psi_0}
\end{equation}
and $\ket{\psi_0}$ is the ground-state of the transfer matrix.
Since the computation and the diagonalization of the reduced density
matrix is required at every iteration of the DMRG algorithm, the
entanglement entropy is easily obtained. $S(L/2)$ is presented
on figure~\ref{Fig3}. Even though, the nature of the phases cannot
be identified solely with the entanglement entropy, the phase
boundaries are much better defined than with more traditional
observables. The phase diagram discussed before is confirmed.

However, the critical phase between the ferromagnetic and paramagnetic
phases cannot be identified anymore, neither in the upper half-plane
nor in the lower one. Moreover, a bump of $S(L/2)$ is observed, as a
yellow thin strip, along a line $r\simeq T-1$ in the paramagnetic phase
of the lower half-plane.
A bump in the specific heat was observed in earlier Monte Carlo simulations
but for larger values of $r$. However, as will be discussed in the following,
the entanglement entropy is hampered by large Finite-Size effects. Therefore,
these bumps may have the same origin. In Ref.~\cite{Baltar}, the bump
in specific heat was interpreted as a phase boundary between the paramagnetic
phase and the anti-ferrimagnetic critical phase. The DMRG data previously
discussed exclude this possibility.


\begin{figure}[ht]
\centering
\psfrag{Ferri}[Bc][Bc][1][1]{\small Ferri}
\psfrag{Ferro}[Bc][Bc][1][1]{\small Ferro}
\psfrag{Para}[Bc][Bc][1][1]{\small Para}
\psfrag{Anti-Ferri}[Bc][Bc][1][1]{\small Anti-Ferri}
\psfrag{B}[Bc][Bc][1][1]{$B$}
\psfrag{A}[Bc][Bc][1][0]{$A$}
\psfrag{r}[Bc][Bc][1][1]{$r$}
\psfrag{T}[Bc][Bc][1][0]{$T$}
\centerline{\includegraphics[width=6.2cm]{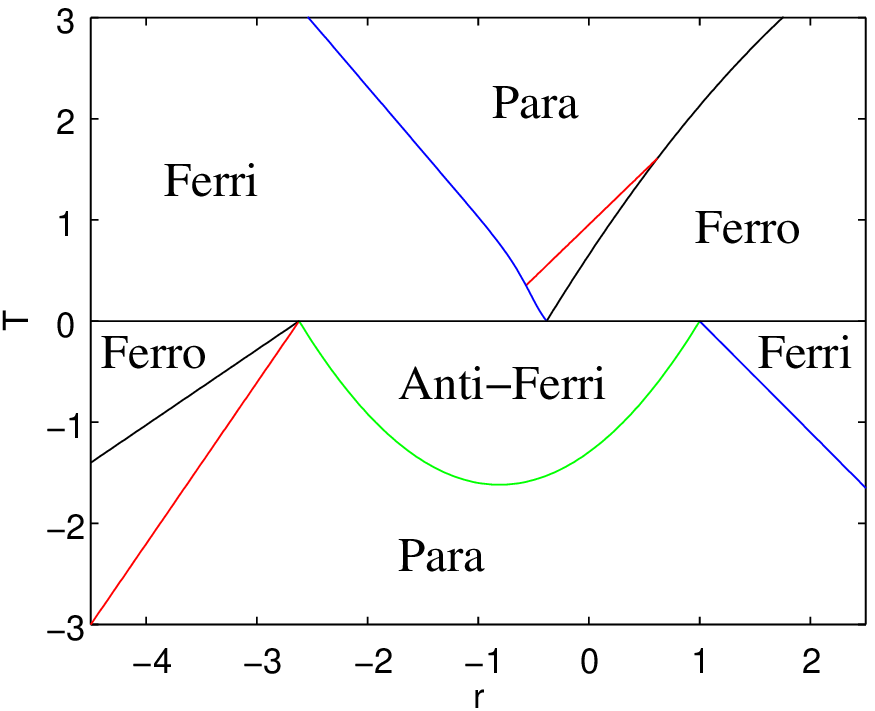}\quad
\includegraphics[width=6.2cm]{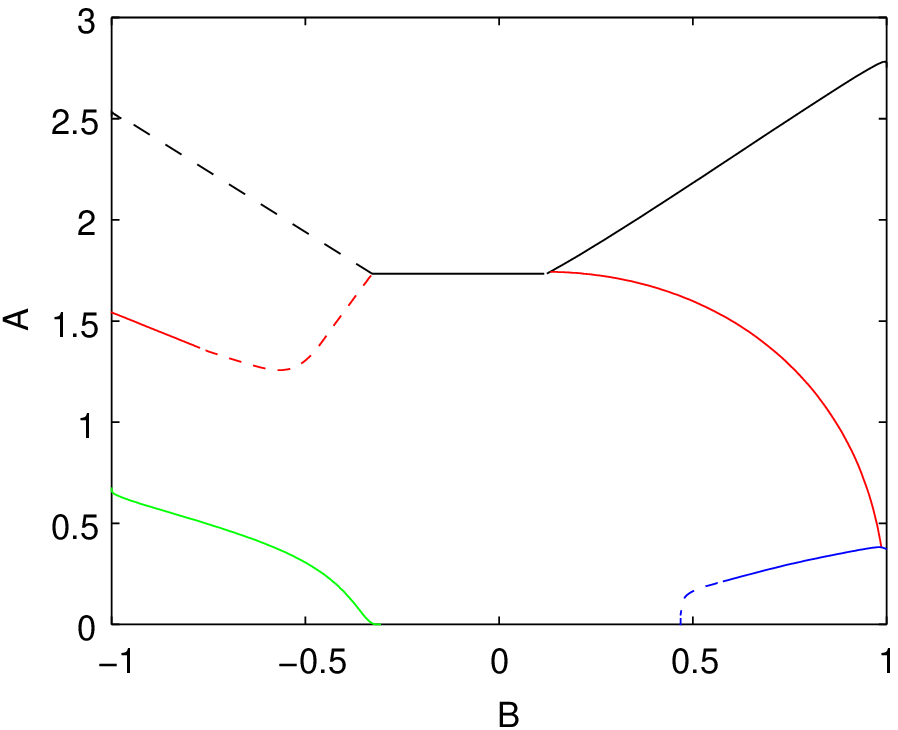}}
\caption{Schematic phase diagram using the variables $(r, T)$ on the left
  and $(A,B)$ on the right.}
\label{FigDenNijs}
\end{figure}

To conclude this section, a schematic phase diagram is proposed on
figure~\ref{FigDenNijs}. On the left, the phase diagram is plotted
in the $(r,T)$ plane. An intermediate critical phase is found between the
ferromagnetic and paramagnetic phases, both at positive and negative $T$.
These two critical phases are bounded by two BKT transitions (black and red
lines). They were not observed in earlier Monte Carlo
simulations~\cite{Baltar}. In the upper half-plane $T>0$, the
intermediate critical phase was considered as a part of the ferrimagnetic
phase. In the lower half-plane, a much larger anti-ferrimagnetic phase,
denoted Phase II, spanning the phase diagram and embracing the critical
phase, was reported. On the right of figure~\ref{FigDenNijs}, the same
phase diagram is plotted in the plane $(A,B)$, as done by den Nijs~\cite{Nijs},
where $A$ and $B$ are related to $x_1$ and $x_2$ (equation~\ref{Defx}) by
\be
A={x_0\over x_1+x_2}\hskip 1cm B={x_1-x_2\over x_1+x_2}
\ee
and $x_0=e^{J_1+J_2}$.
The map $(r,T)\rightarrow (A,B)$ is highly non-linear. Some regions of the phase
diagram are exponentially shrunk, while others are exponentially expanded.
Therefore, while the phase diagram in the $(r,T)$ plane reproduces closely the
numerical data, it may not be the case of the same diagram in the $(A,B)$
plane. Note that the line $T=0$ corresponds to $B=\pm 1$.
Furthermore, only the bold lines were obtained from the numerical data.
The dashed lines are only a guess.
As predicted by den Nijs, the anti-ferrimagnetic phase appears at the
bottom-left corner of the phase diagram while the two ferrimagnetic phases
are mapped onto the bottom-right corner. The intermediate critical phase
observed in the $T>0$ half-plane occupies a large region on the right of
the diagram. In contrast, the critical phase of the $T<0$ half-plane is
mapped onto a thin region lying on the left $B=-1$ axis. The dashed lines
plotted on the phase diagram is a guess based on den Nijs' predictions.
It implies that, like in the upper half-plane, the two BKT transitions
meets at a point with $r<-4.5$, i.e. outside of the range of couplings
that were studied. The phase diagram is essentially in agreement with
den Nijs' scenario (b), discussed at the end of section~\ref{Intro}.
However, this agreement is not complete. In the
upper-half plane, the BKT transition between the paramagnetic and critical
phase (red line) was predicted to terminate on the $T=0$ axis at the
contact point $r\simeq -0.382$ between the ferromagnetic and ferrimagnetic
phases. In contrast, we observe a finite contact between the ferrimagnetic
and critical phase. This situation, that would correspond to scenario (c),
is not compatible with figure 1 of Ref.~\cite{Nijs} but with the
predicted phase diagram for the 7-state $\mathbb{Z}_q$ model (figure 4).
In contrast, in the quarter $r<0$ and $T<0$, the contour lines of all
observables reach the $T=0$ axis at the same point $r\simeq -2.618$. As
a consequence, one may infer that the two BKT transition lines meet
at this point of the $T=0$ axis, meaning that the critical phase does not
have any common boundary with the anti-ferrimagnetic phase, as expected in
den Nijs' scenario (b). However, a more direct and precise estimation of
the location of the BKT transition line is needed to confirm this statement.

\subsection{Finite-size scaling of entanglement entropy and free energy}
In non-critical phases, local degrees of freedom are expected to be
entangled over a distance of the order of the correlation length
$\xi$. Therefore, the entanglement entropy is finite, $S\sim \ln\xi$,
and does not depend on the lattice size $L$ as long as $L>\xi$. At
critical points or in critical phases, the correlation length
becomes infinite and the entanglement entropy grows with the lattice
size as $S\sim \kappa\ln L$. In conformal field theories (CFT), the prefactor
$\kappa$ is universal and proportional to the central charge $c$~\cite{%
Vidal,Cardy}. For periodic boundary conditions, $\kappa=c/3$ while
$\kappa=c/6$ for open ones.

\begin{figure}[ht]
\psfrag{r=J2/J1}[Bc][Bc][1][1]{$r$}
\psfrag{S}[Bc][Bc][1][0]{$S(L/2)$}
\centerline{
  \includegraphics[width=7cm]{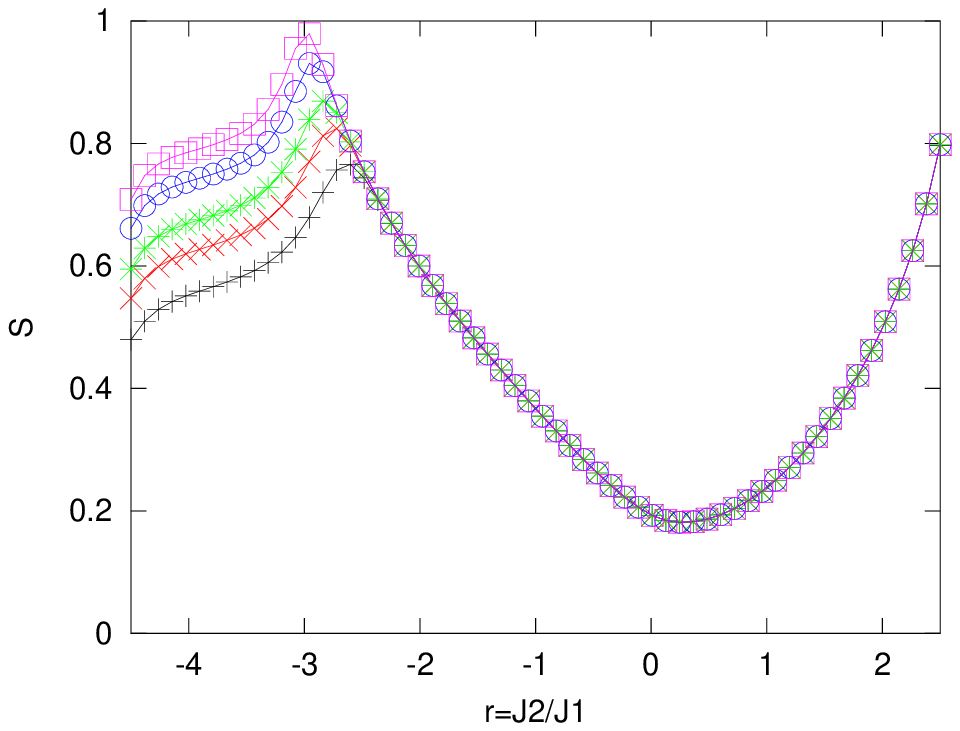}
  \includegraphics[width=7cm]{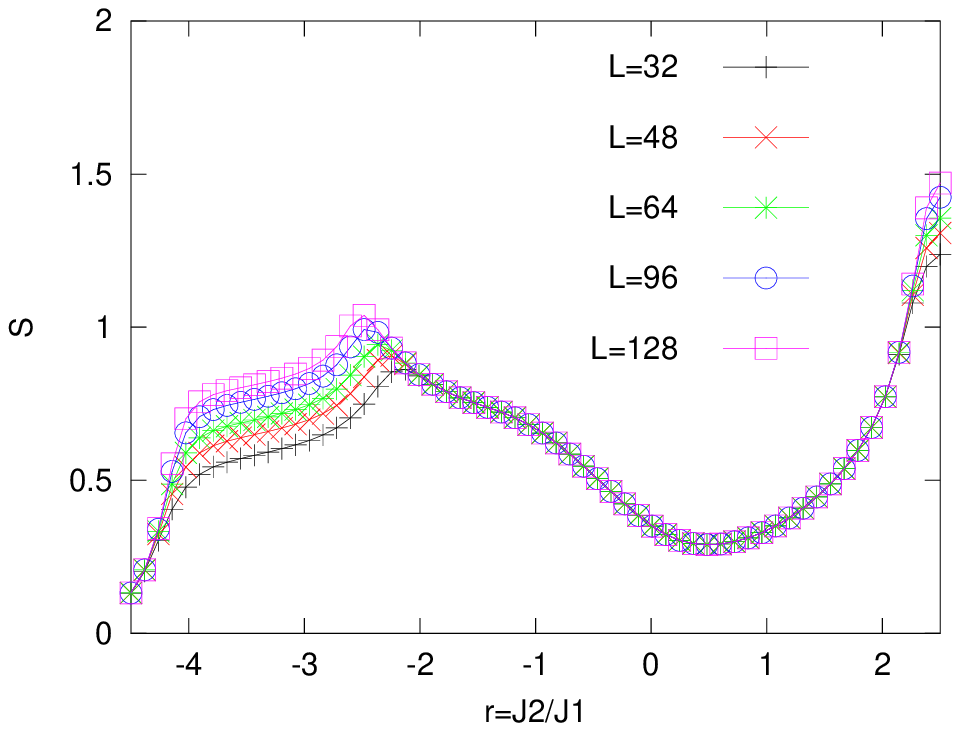}}
\centerline{
  \includegraphics[width=7cm]{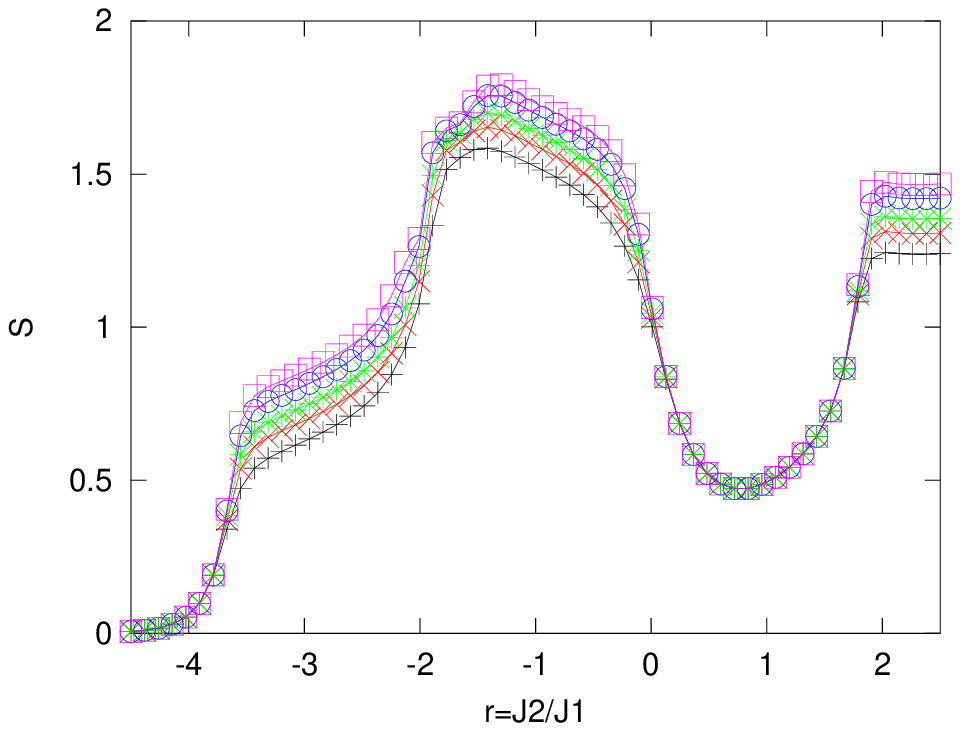}
  \includegraphics[width=7cm]{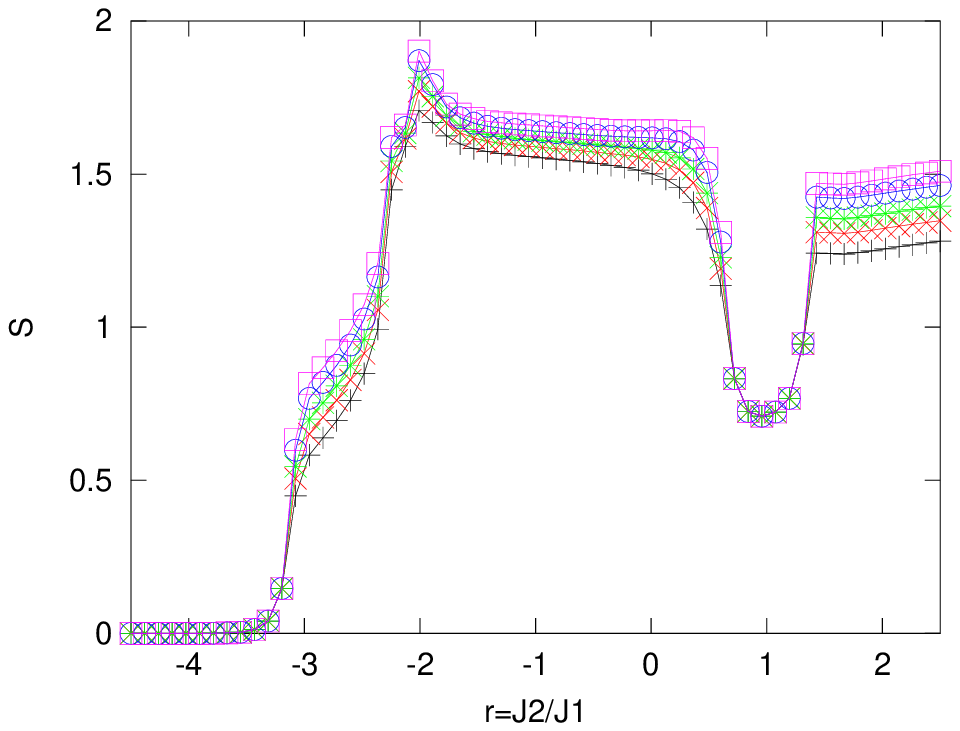}}
\caption{Entanglement entropy versus $r=J_2/J_1$. The different graphs
  correspond to $T=1/J_1\simeq -1.983$ (top left), $-1.475$, $-0.966$ and
  $-0.458$ (bottom right).}
\label{Fig6}
\end{figure}

On figure~\ref{Fig6}, the entanglement entropy $S(L/2)$ is plotted
for four different couplings $J_1=1/T$ versus $r=J_2/J_1$. The curves
for $L=128$ correspond to horizontal sections of figure~\ref{Fig3}.
One can clearly distinguish regions where the curves corresponding
to different lattice sizes fall on top of each other. These regions of
$L$-independent entanglement entropy are therefore non-critical.
In contrast, three critical phases are observed, in agreement with
the discussion of the previous section. At $T\simeq -1.983$, only the
critical phase that we conjectured to exist between the ferromagnetic
and paramagnetic phases, is observed. This critical phase is bounded
by a peak whose location corresponds precisely to the yellow stripe
observed on figure~\ref{Fig3}. However, this peak is strongly shifted
to smaller values of $r$ as the lattice size is increased. Entanglement
entropy being a non-local quantity, it is indeed more sensitive to
boundaries than local observables. Therefore, while entanglement entropy
was useful to enlighten the critical nature of this phase, its boundaries
are probably more accurately given by magnetization-magnetization
autocorrelation $A_1$ for example. As $T=1/J_1$ is decreased, a second
critical phase is observed on figure~\ref{Fig6}, corresponding
to the anti-ferrimagnetic phase. Finally, at $T\simeq -0.458$, the
ferrimagnetic critical phase is observed on the right of the figure.

\begin{figure}[ht]
\psfrag{L}[Bc][Bc][1][1]{$L$}
\psfrag{1/L2}[Bc][Bc][1][1]{$1/L^2$}
\psfrag{S}[Bc][Bc][1][0]{$S(L/2)$}
\psfrag{F}[Bc][Bc][1][0]{$f-f_\infty-f_s/L$}
\centerline{\includegraphics[width=6.6cm]{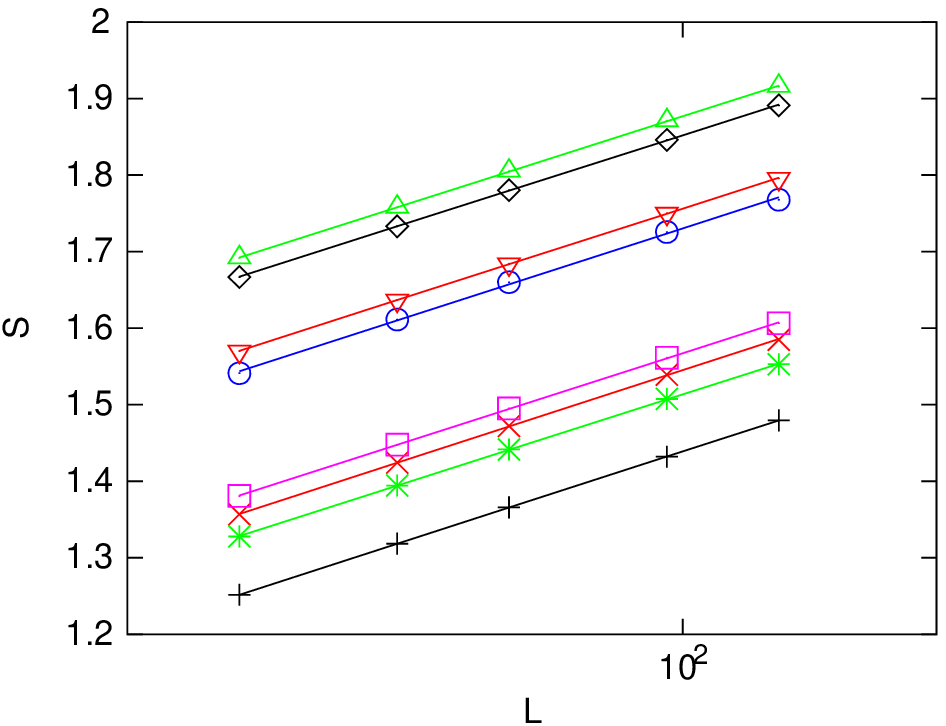}
\quad\quad\includegraphics[width=7.4cm]{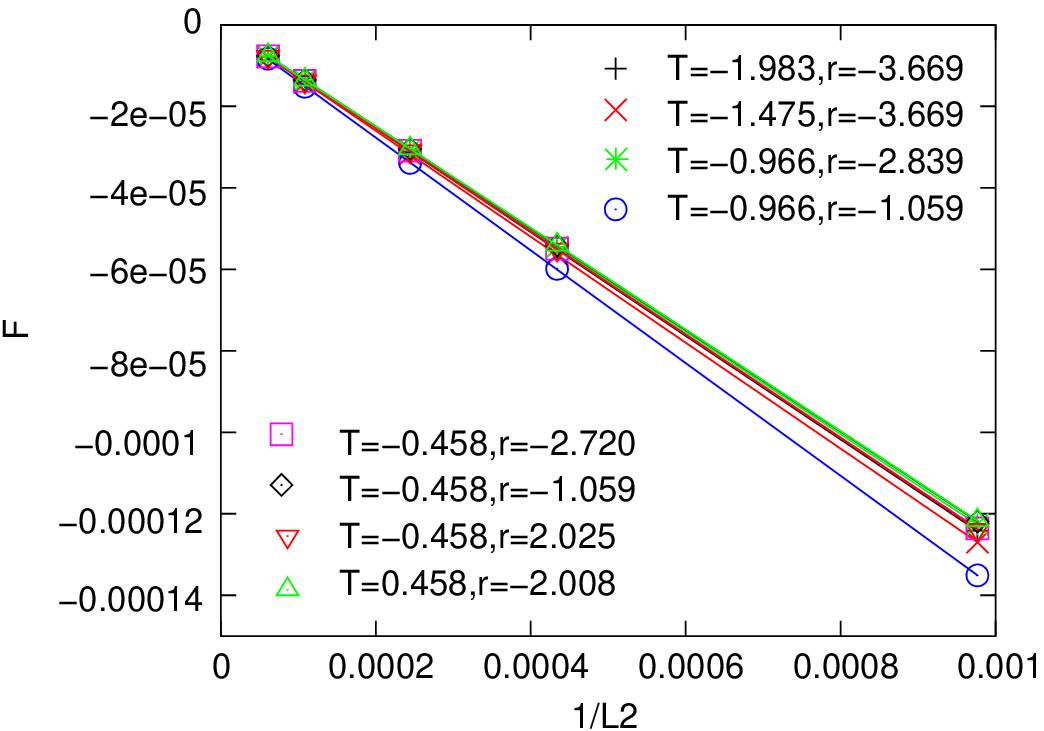}}
\caption{On the left, entanglement entropy $S(L/2)$ versus $L$
  in a semi-logarithmic scale at different points of the phase diagram.
  The lines are linear fits. On the right, free energy density after
  the removal of the thermodynamic limit $f_\infty$ and the surface
  free energy $f_s$. The symbols are the same on the two figures.}
\label{FigS}
\end{figure}

On figure~\ref{FigS}, the entanglement entropy $S(L/2)$ is plotted
versus $\ln L$ at different points of the phase diagram, all of them
belonging to critical phases. As expected, the behavior is very nicely
linear with a slope in the range $0.1620(7)$ ($T\simeq -0.458$, $r\simeq
-2.839$) to $0.1649(7)$ ($T\simeq -1.475$, $r\simeq -3.669$). These
values are slightly below but close to the conformal prediction
$\kappa=c/6$ with a central charge $c=1$. This result is consistent
with the existence of a mapping of the ground state in the ferrimagnetic
and anti-ferrimagnetic critical phases onto a Solid-On-Solid model, i.e.
a Gaussian theory, as discussed in the second section. Note that the
data presented on figure~\ref{FigS} result from additional numerical
calculations with open boundary conditions, i.e. without boundary
magnetic fields, as required by conformal theory. In presence of
boundary fields, the prefactor $\kappa$ would be slightly smaller
in the anti-ferrimagnetic phase.
\\

The same information can be extracted from the Finite-Size Scaling of
the free energy density. In presence of open boundary conditions,
the latter is predicted by CFT to behave as
\begin{equation}
  f(L)=f_\infty+{f_s\over L}+{\kappa'\over L^2}+{\cal O}(L^{-3})
\end{equation}
where $f_s$ is a surface free energy density and $\kappa'=\pi c/24$.
This behavior is clearly observed and a parabolic fit gives the three
parameters $f_\infty$, $f_s$ and $\kappa'$ with a good accuracy.
On figure~\ref{FigS}, the quantity $f(L)-f_\infty-f_s/L$, where $f(L)$
corresponds to the numerical data while $f_\infty$ and $f_s$ result from
the fit, is plotted versus $1/L^2$. The nice linear behavior that is
observed shows that the higher-order terms in $1/L$ can be safely
neglected. $\kappa'$ is found in the range $-0.1246$ ($T\simeq -0.458$,
$r\simeq -2.008$) to $-0.1383$ ($T\simeq -0.966$, $r\simeq -1.059$).
The corresponding central charges are in the range $0.952-1.056$.

\subsection{Berezinskii-Kosterlitz-Thouless transitions in the AF regime}
In this section, the location of the BKT transition between the critical
phase and the paramagnetic phase is more precisely determined.
Despite the lack of spontaneous symmetry breaking in the critical phase,
an order parameter, the helicity modulus $\Upsilon$, can be
defined~\cite{Fisher}. Twisting boundary conditions are imposed to the
system to induce a spin wave. The helicity modulus is related to the
variation of free energy density as
 \begin{equation}
  \Upsilon=L^2\left({\partial^2 f\over\partial \Delta^2}\right)_{\Delta=0}
 \end{equation}
where $\Delta$ is the angle difference imposed to spins at
the left and right boundaries. In the XY model, whose Hamiltonian
is $-J\sum_{(i,j)}\cos(\theta_i-\theta_j)$, the helicity modulus
is expected to display a jump equal to the stiffness $J$
at the BKT transition in the thermodynamic limit.
In practise, this quantity is computed numerically by imposing
different orientations of the magnetic field at the two boundaries.
In the case of the $\mathbb{Z}(q)$ model, the Hamiltonian
(\ref{BoundaryH}) is replaced by
\begin{equation}
  -\beta H_{\rm BC}=\cos{2\pi\over q}\sigma_1
  +\cos{2\pi\over q}(\sigma_L-\Delta).\label{BoundaryH2b}
\end{equation}
where $\Delta$ is a now discrete quantity. The free energy density
$f(\Delta)$ is computed for $\Delta=0$ and $1$ and the helicity modulus
is estimated as
\begin{equation}
  \Upsilon={2L^2\over (2\pi/q)^2}\big(f(0)-f(1)\big).
  \end{equation}
This procedure has been shown to give the correct helicity modulus in the
case of the 5-state clock model, i.e. $J_2=0$ here~\cite{ChatelainZq}.
It should therefore extend to the BKT transition between the critical
phase and the ferromagnetic and paramagnetic phases.
In the case of the ferrimagnetic and anti-ferrimagnetic phases, the boundary
fields introduce a frustration at both boundaries. The contribution of
the two additional surface free energies should however cancel in the
definition of the helicity modulus.
\\

\begin{figure}[ht]
\psfrag{r=J2/J1}[Bc][Bc][1][1]{$r$}
\psfrag{m}[Bc][Bc][1][0]{$\Upsilon$}
\centerline{
  \includegraphics[width=7cm]{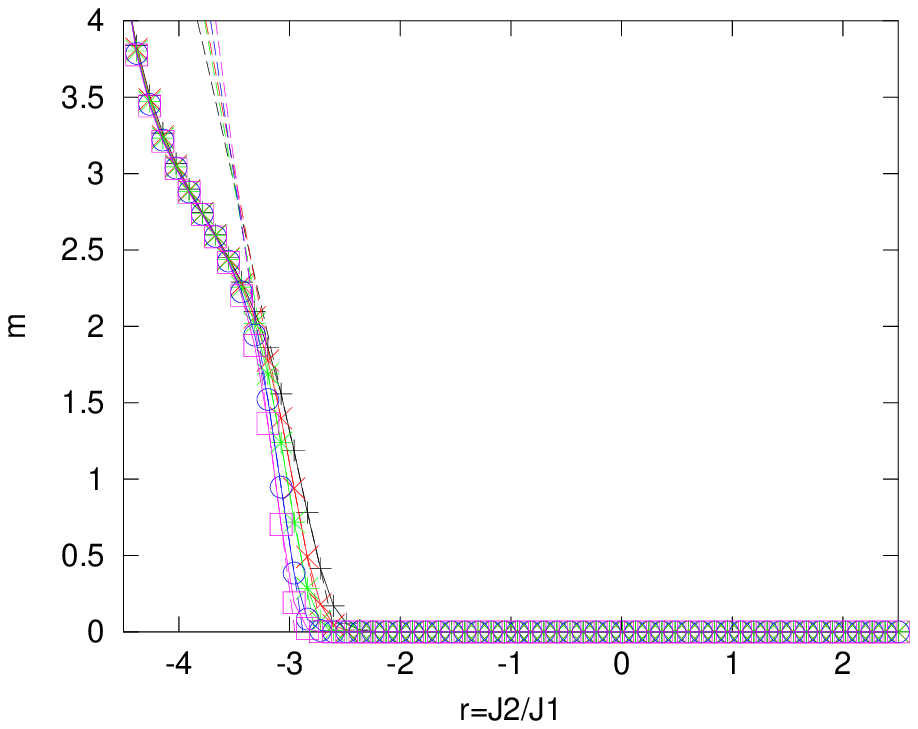}
  \includegraphics[width=7cm]{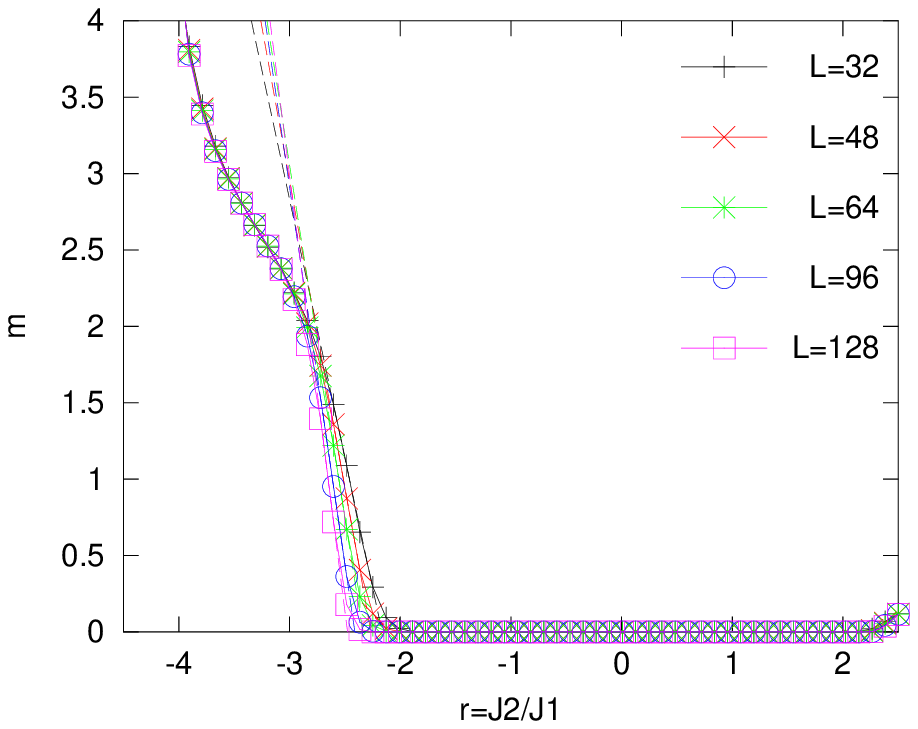}}
\centerline{
  \includegraphics[width=7cm]{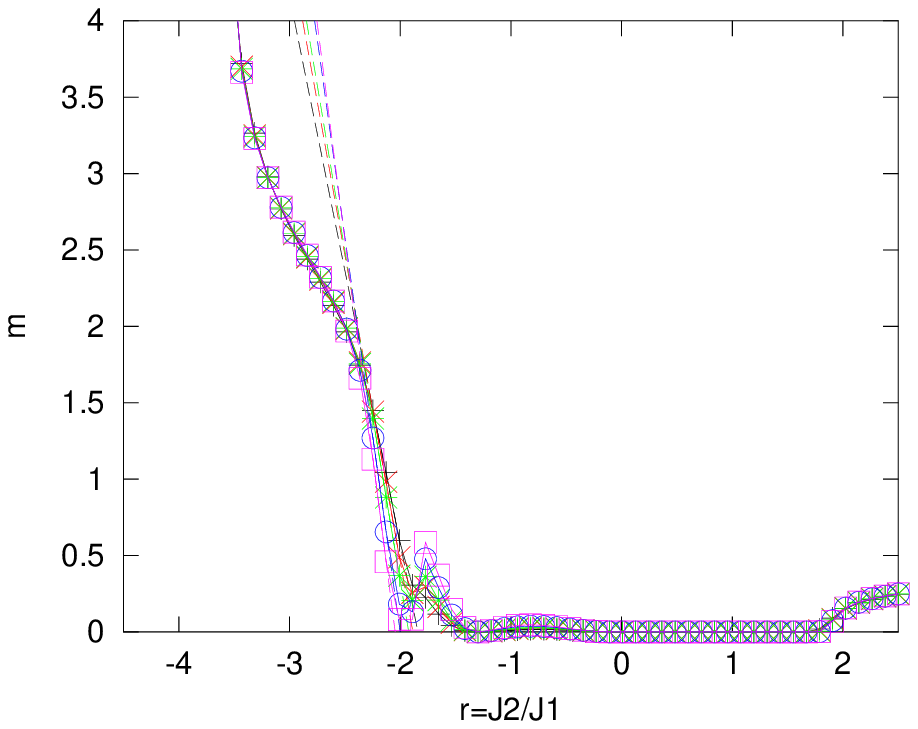}
  \includegraphics[width=7cm]{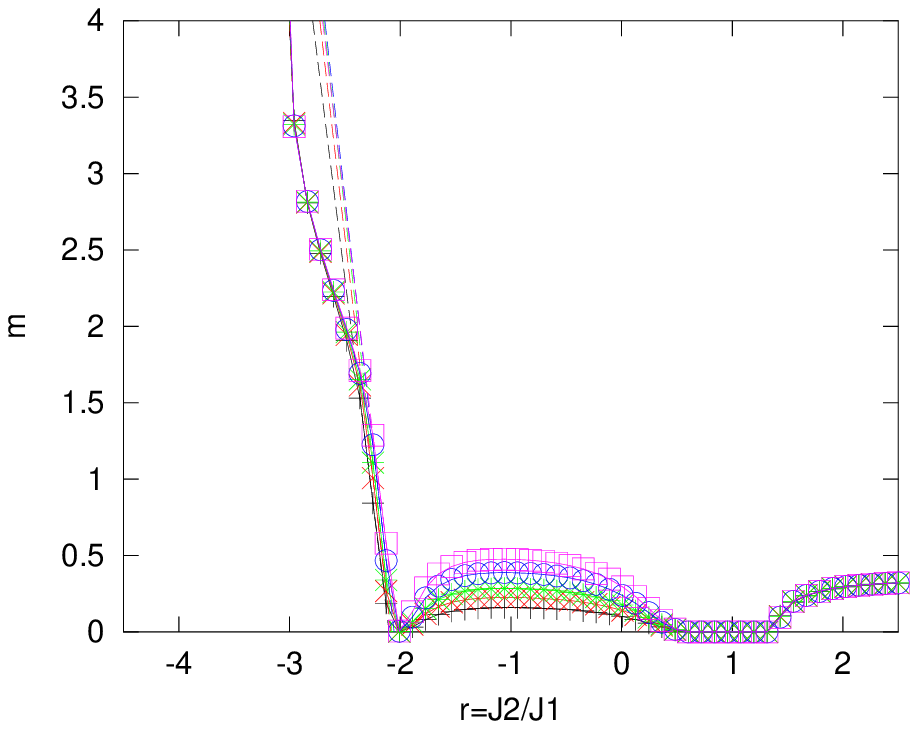}}
\caption{Helicity modulus $\Upsilon$ versus $r=J_2/J_1$. The different graphs
  correspond to $T=1/J_1\simeq -1.983$ (top left), $-1.475$, $-0.966$ and
  $-0.458$ (bottom right). The different curves correspond to different
  lattice sizes as indicated in the legend. The dashed lines are linear
  fits of the helicity modulus in the region of sharpest variation.}
\label{Fig8}
\end{figure}

The helicity modulus $\Upsilon$ is plotted on figure~\ref{Fig8} versus
$r=J_2/J_1$ for four different temperatures. For $T=1/J_1\simeq -1.983$ and
$-1.475$, the curves are typical of a model undergoing a
BKT transition and very similar to those observed for the 5-state
clock model~\cite{ChatelainZq}. As the lattice
size $L$ is increased, the curves become steeper and steeper, as expected
for a quantity displaying a jump in the thermodynamic limit. To estimate
the location of the transition, the helicity modulus is fitted linearly
in the region of sharpest variation. Those fits are represented on
figure~\ref{Fig8}. As expected, the slope increases with the lattice size.
The intercept with the $x$-axis is then extrapolated in the limit
$L\rightarrow +\infty$ using a linear fit with $1/L$. The BKT transition
points are estimated to be located at $r_{\rm BKT}=-3.29(21)$ for $T\simeq
-1.983$, $-2.77(13)$ for $T\simeq -1.475$, $-2.24(28)$ for $T\simeq -0.966$,
and $-1.97(12)$ for $T\simeq -0.458$.

For $T\simeq -0.458$, the anti-ferrimagnetic and ferrimagnetic critical
phases appear as a bump in the helicity modulus $\Upsilon$. However, we
do not know how to interpret this to extract useful informations since
the topological transition is not of BKT-type in these cases.
\\

The analysis of helicity modulus leads to estimates of the value of $r$
at the BKT transition larger than those reported in the previous section
\ref{Sec1}. Since the helicity modulus is an order parameter of the BKT
transition, while quantities studied in section~\ref{Sec1} are not, and
since an extrapolation to the thermodynamic limit was moreover performed,
the estimates of this section are believed to be more reliable. There is
an important consequence: the estimates of $r_{\rm BKT}$ for $T<-0.966$
are larger than $-2.618$ ($-1.97(12)$ for $T\simeq -0.458$ for instance).
It means that the BKT transition line presumably does not terminate on
the $T=0$ axis at the point $r\simeq -2.618$ where ferromagnetic and
anti-ferrimagnetic phases meet but on the boundary of the anti-ferrimagnetic
phase. The critical phase and the anti-ferrimagnetic phase have therefore a
finite common boundary, like in the $T>0$ half-plane of the phase diagram. As
a consequence, for both $T>0$ and $T<0$, the phase diagram is compatible
with den Nijs' prediction for the 7-state $\mathbb{Z}_q$ model (figure 4
of \cite{Nijs}).

\subsection{Autocorrelation functions}

\begin{figure}[ht]
\psfrag{t}[Bc][Bc][1][1]{$t$}
\psfrag{Ce(t)}[Bc][Bc][1][1]{$C_{ee}(t)$}
\psfrag{Cm(t)}[Bc][Bc][1][0]{$C_{\sigma\sigma}(t)$}
\centerline{\includegraphics[height=5.25cm]{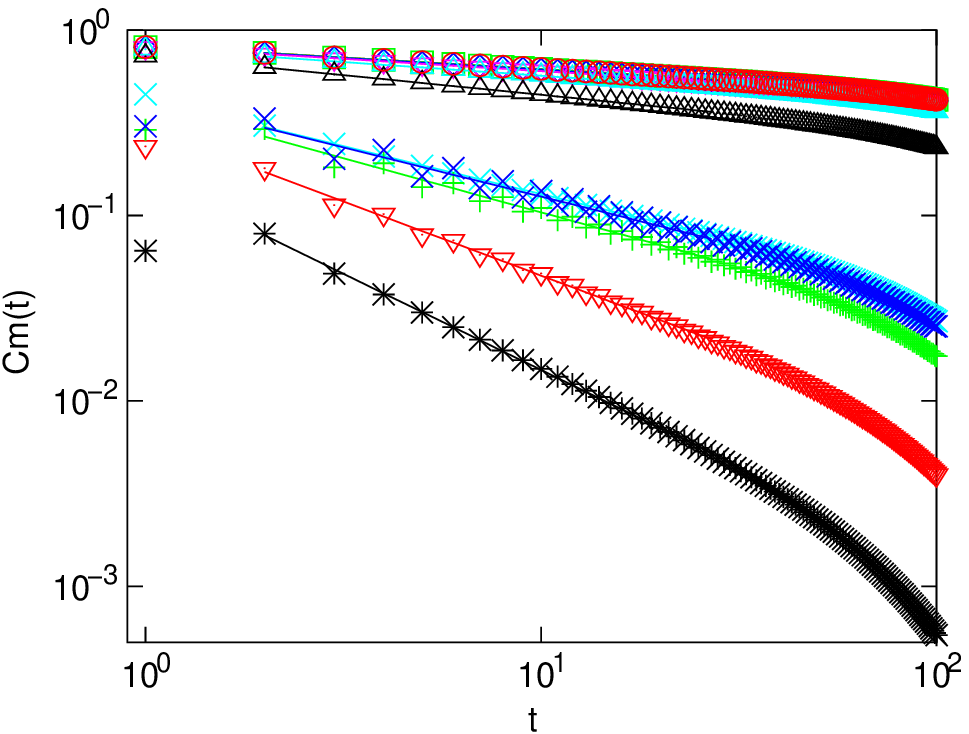}\quad
\includegraphics[height=5.25cm]{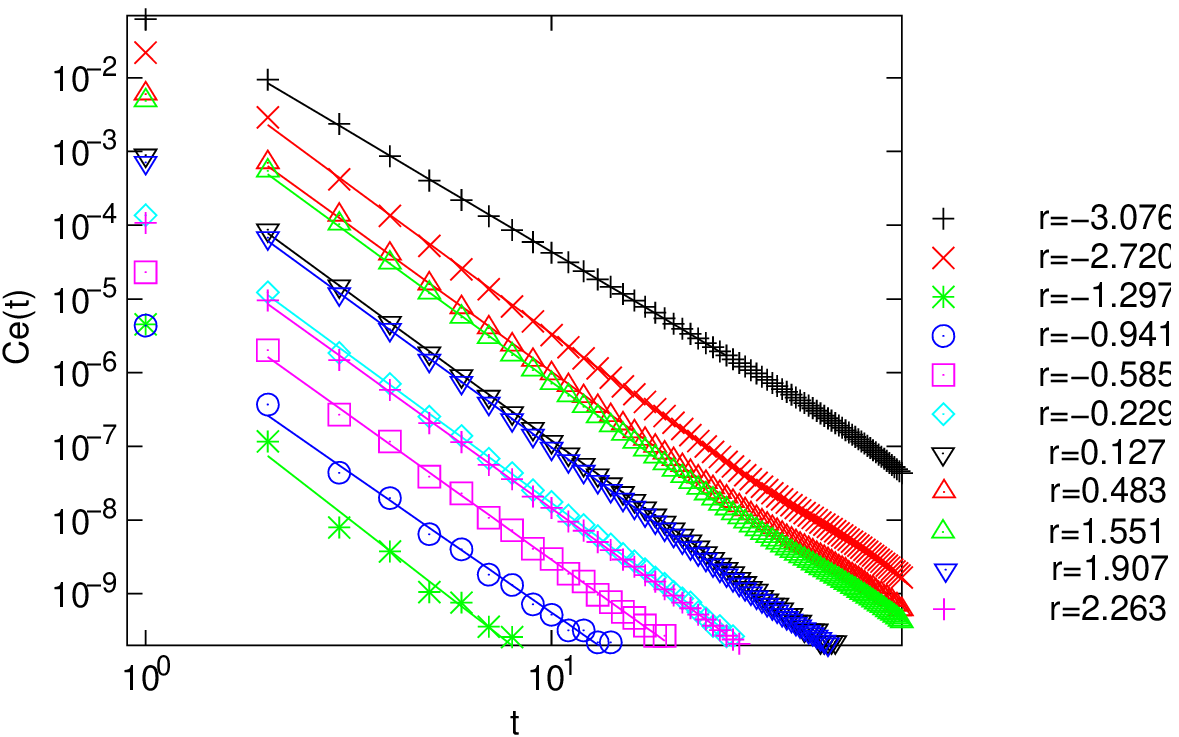}}
\caption{On the left, absolute value of the spin-spin correlation functions
at different values of $r$ and at $T=1/J_1\simeq -0.458$. The lines are
log-log fits. On the right, energy-energy correlation functions at the
same values of $r$ and $T$.}
\label{FigC}
\end{figure}

In this section, the correlation functions are shown to decay algebraically
in all critical phases and an associated critical exponent is estimated.
On figure~\ref{FigC}, the absolute value of spin-spin autocorrelation functions
  \begin{equation}
    C_{\sigma\sigma}(t)=|\langle \cos{2\pi\over q}\big[\sigma_{L/2}(t)
    -\sigma_{L/2}(0)\big]\rangle|
  \end{equation}
and energy-energy autocorrelation functions
  \begin{equation}
    C_{ee}(t)=\langle e(t)e(0)\rangle-\langle e\rangle^2
  \end{equation}
with
  \begin{equation}
    e(t)=\cos{2\pi\over q}(\sigma_{L/2}(t)-\sigma_{L/2+1}(t))
  \end{equation}
are plotted for $T\simeq -0.458$ and for values of $r$ that correspond
to points of the phase diagram in critical phases. The absolute value
was introduced because in the anti-ferrimagnetic phase, the spin-spin
correlation function is oscillating. For all points considered, a
nice algebraic decay is observed, at least for small times $t\lesssim 30$
due to the finite size of the system. For larger times, an exponential
decay is recovered.

\begin{figure}[ht]
\psfrag{r}[Bc][Bc][1][1]{$r$}
\psfrag{eta}[Bc][Bc][1][1]{$\eta$}
\centerline{\includegraphics[width=7cm]{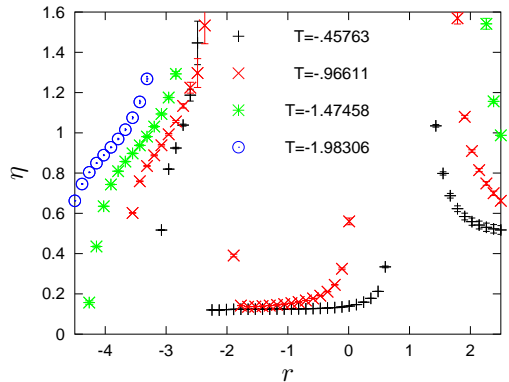}}
\caption{Critical exponent $\eta$, computed from a log-log fit of
spin-spin autocorrelation functions, versus $r$. The different
symbols correspond to different values of $T$. To exclude the
points of the phase diagram in non-critical phases (ferromagnetic
and paramagnetic), two criterions have been used: the entanglement
entropy $S(L/2)$ was required to be larger than $0.7$ and $r$
to be outside the range $[r_{\rm BKT}(T);-2]$ where $r_{\rm BKT}(T)$ is
the location of the BKT transition as given by the jump of the
helicity modulus.}
\label{FigEta}
\end{figure}

The critical exponent $\eta$ was computed by a log-log fit of spin-spin
autocorrelation functions. As can be seen on figure~\ref{FigEta}, three
distinct behaviors are observed. First, in the critical phase lying
between the ferromagnetic and paramagnetic phases, the exponent $\eta$
starts with a small value at the first BKT transition, i.e. at the boundary
of the ferromagnetic phase, and then grows rapidly as $r$ is increased.
The rapid variation of $\eta$ with $r$ makes difficult the estimation
of its value along the second BKT transition line, i.e. between the
critical and paramagnetic phase. The exponent takes a value in the range
$1.3-1.5$ and may be constant along the BKT transition line. In the
anti-ferrimagnetic phase, the exponent $\eta$ displays a plateau, clearly
seen for $T\simeq -0.457$. The estimate vary from $0.1232$ at $r\simeq -1.9$
to $0.1262$ at $r\simeq -0.7$. It is tempting to conjecture that the
exponent $\eta$ takes the rational value $1/16$, i.e. half of the exponent
of the Ising model, in all points of the anti-ferrimagnetic phase. However,
at the right boundary of the anti-ferrimagnetic phase an increase of $\eta$
is observed. It may not be physical but due to finite size effects. For
$T\lesssim -1$, there is no anti-ferrimagnetic phase anymore. Finally, in
the ferrimagnetic phase, the exponent $\eta$ decreases with $r$. It is not
possible to say whether $\eta$ takes the same value along the topological
transition line or not.
\\

An exponent $\eta_e=2x_e$ have also been computed by a log-log fit
of the energy-energy autocorrelation functions. However, it is much
noisier than $\eta$ and we will not draw any conclusion about it.

\section{Conclusions}
The phase diagram of the 5-state ${\mathbb Z}(q)$ model has been
determined numerically using Density Matrix Renormalization Group
and compared to earlier Monte Carlo simulations. In the quadrant
$T=1/J_1<0$ and $r=J_2/J_1<0$ where the two interactions are frustrated,
a new phase diagram is proposed. The critical anti-ferrimagnetic
phase has a finite extension in the $(r,T)$ plane and does not share
any boundary with the ferromagnetic phase, apart from one point in
the limit $T\rightarrow 0^-$. At the boundary of the ferromagnetic phase
lies a critical phase. Similarly to what happen when $T,r>0$, the system
undergoes two successive BKT transitions when going from the paramagnetic
phase to the ferromagnetic one. This phase diagram is supported by the
behavior of the entanglement entropy, the helicity modulus, the
critical exponent $\eta$, and contradicts earlier Monte Carlo simulations.
\\

\begin{figure}[ht]
\centering
\psfrag{Ferri}[Bc][Bc][1][1]{\small Ferri}
\psfrag{Ferro}[Bc][Bc][1][1]{\small Ferro}
\psfrag{Para}[Bc][Bc][1][1]{\small Para}
\psfrag{Anti-Ferri}[Bc][Bc][1][1]{\small Anti-Ferri}
\psfrag{B}[Bc][Bc][1][1]{$B$}
\psfrag{A}[Bc][Bc][1][0]{$A$}
\psfrag{r}[Bc][Bc][1][1]{$r$}
\psfrag{T}[Bc][Bc][1][0]{$T$}
\centerline{\includegraphics[width=6.2cm]{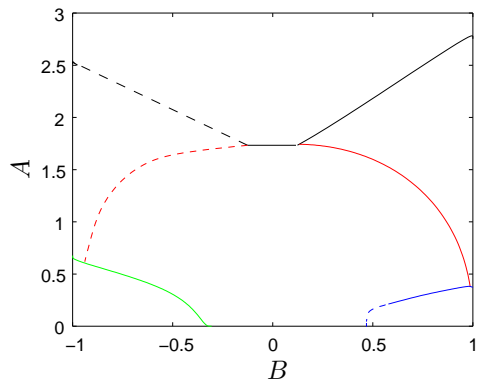}}
\caption{Schematic phase diagram using the variables $(A,B)$ defined
  by den Nijs in Ref.~\cite{Nijs}.}
\label{FigDenNijsB}
\end{figure}

In the $T>0$ half plane, the intermediate critical phase has a common
boundary with the ferrimagnetic phase, which invalidates the scenario (b)
proposed in~\cite{Nijs} and considered to be ``the more likely''. However,
the phase diagram proposed for the ${\mathbb{Z}}_7$ model in the same
reference would be in agreement with the data. In the $T<0$ half plane,
the situation is more confuse since the behavior of magnetization or two-site
correlations could be interpreted as the absence of connection between the
intermediate critical phase and the anti-ferrimagnetic phase. However, the
helicity modulus, which is the order parameter of the BKT transition,
indicates the existence of a common boundary between the intermediate
critical phase and the anti-ferrimagnetic phase. A schematic phase diagram
corresponding to these data is represented on figure \ref{FigDenNijsB}.
The phase diagram is finally in agreement with the one proposed for the
${\mathbb{Z}}_7$ model in Ref.~\cite{Nijs}.
\\

These numerical results call for a reconsideration of the mechanism
underlying the phase transitions in the ${\mathbb{Z}}_5$ model. More
complex composite objects than those considered in Ref.~\cite{Nijs}
in the case $q=5$ should already be present in the low-temperature phases
and have a relevant contribution in the transitions to the paramagnetic
phase.

\section*{Acknowledgements}
The author is indebted to the anonymous referees for useful comments and
in particular, for having pointed out that the phase diagram is in agreement
with figure 4 of Ref.~\cite{Nijs}.


\section*{References}
\begin{thebibliography}{10}
\bibitem{Berezinski}
V.~L. Berezinskii. {\sl Zh. Eksp. Teor. Fiz.}, {\bf 61} 1144, (1971).

\bibitem{Kosterlitz}
J.~M. Kosterlitz and D.~J. Thouless
{\sl J. Phys. C: Solid State Physics}, {\bf 6} 1181, (1973).

\bibitem{Villain}
  J. Villain {\sl J. Phys. C: Solid State Physics} {\bf 10}
  1717 (1977)

\bibitem{Blote}
  H.W.J. Bl\"ote, M.P. Nightingale, X.N. Wu and A. Hoogland
  {\sl Phys. Rev. B} {\bf 43} {8751} (1991) ;
  H.W.J. Bl\"ote and M.P. Nightingale {\sl Phys. Rev. B}
  {\bf 47} {15046} (1993).

\bibitem{Foster}
D.P. Foster, C. G\'erard and I. Puha {\sl J. Phys. A} {\bf 34} {5183} (2001)

\bibitem{Villain2}
J. Villain {\sl J. Phys. C} {\bf 10} {4793} (1977)

\bibitem{Quartin}
M. Quartin and S.L.A de Queiroz {\sl J. Phys. A: Math. Gen.}
{\bf 36} 951 (2003)

\bibitem{Ding}
C. Ding, H.W.J. Bl\"ote and Y. Deng {\tt arXiv:1508:04538} (2015)

\bibitem{Selke}
  W. Selke, {\sl Phys. Rep.} {\bf 170} 213 (1988).

\bibitem{Nijs2}
  M. den Nijs {\sl Phys. Rev.} {\bf 31}, 266 (1985).

\bibitem{LeeGrinstein}
  D.H. Lee and G. Grinstein {\sl Phys. Rev. Lett.} {\bf 55}, 541 (1985).

\bibitem{Dian}
M. Dian and R. Hlubina {\sl Phys. Rev. B} {\bf 84}, 224420 (2011).

\bibitem{Poderoso}
  F.C. Poderoso, J.J. Arenzon and Y. Levin {\sl Phys. Rev. Lett.}
  {\bf 106}, 067202 (2011).

\bibitem{Canova}
  G.A. Canova, Y. Levin and J.J. Arenzon {\sl Phys. Rev. E} {\bf 89}
  012126 (2014).

\bibitem{Jose}
  J.V. Jos\'e, L.P. Kadanoff, S. Kirkpatrick, and D.R. Nelson
  {\sl Phys. Rev. B} {\bf 16}, 1217 (1977).

\bibitem{Nijs}
  M. den. Nijs {\sl Phys. Rev. B} {\bf 31}, 266 (1985).

\bibitem{Baltar}
  V.L.V. Baltar, G.M. Carneiro, M.E. Pol, and N. Zagury
  {\sl J. Phys. A: Math. and Gen.} {\bf 18} 2017 (1985).

\bibitem{Rouidi}
  K. Rouidi and Y. Leroyer {\sl Phys. Rev. B} {\bf 45} 1013 (1992).
  
\bibitem{Potts}
R.~B. Potts. {\sl Mathematical Proceedings of the Cambridge
Philosophical Society}, {\bf 48} 106, (1952).

\bibitem{Alcaraz}
F.C. Alcaraz and R. Koberle {\sl J. Phys. A: Math. Gen.}
{\bf 13} L153 (1980).


\bibitem{Fateev}
  V.A. Fateev, and A.B. Zamolodchikov {\sl Phys. Lett. A}
  {\bf 92}, 37 (1982).

\bibitem{Lieb}
E. H. Lieb {\sl Phys. Rev.} {\bf 162} 162 (1967).
  
\bibitem{White}
S.~R. White {\sl Phys. Rev. Lett.}, {\bf 69} 2863, (1992) ;
S.~R. White {\sl Phys. Rev. B}, {\bf 48} 10345, (1993).

\bibitem{Schollwock}
U.~Schollw\"ock {\sl Rev. Mod. Phys.}, {\bf 77} 259, (2005).

\bibitem{Amico}
  L. Amico, R. Fazio, A. Osterloh, and V. Vedral (2008)
  {\sl Rev. Mod. Phys.} {\bf 80}, 517.
  
\bibitem{Vidal}
G. Vidal, J.I. Latorre, E. Rico, and A. Kitaev (2003)
{\sl Phys. Rev. Lett.} {\bf 90}, 227902.

\bibitem{Cardy}
P. Calabrese and J. Cardy (2004) {\sl J. Stat. Mech.: Theory Exp.} P06002.

\bibitem{Fisher}
M.~E. Fisher, M.~N. Barber, and D.~Jasnow.
{\sl Phys. Rev. A}, {\bf 8} 1111, (1973).

\bibitem{ChatelainZq}
C. Chatelain {\sl J. Stat. Mech.: Theory Exp.} P11022 (2014).

\end{thebibliography}
\end{document}